\documentclass[%
 reprint,
 amsmath,amssymb,
 aps,
]{revtex4-1}

\usepackage[dvipdfmx]{graphicx} % Include figure files
\usepackage{dcolumn}% Align table columns on decimal point
\usepackage{bm}% bold math
\usepackage[font=small,format=plain,labelformat=simple,labelsep=period,justification=raggedright]{caption}
\usepackage{amsmath}
\usepackage{here}
\usepackage{natbib}
\usepackage[dvipdfmx]{color}
\usepackage{url}

\begin{document}

\preprint{APS/123-QED}

\title{Comparison of escalator strategies in models using \\ a modified totally asymmetric simple exclusion process}% Force line breaks with \\

\author{Hiroki Yamamoto$^{1}$}
\email{h18m1140@hirosaki-u.ac.jp}
\thanks{}%
\author{Daichi Yanagisawa$^{2,3}$}%
\author{Katsuhiro Nishinari$^{2,3}$}%
\affiliation{%
$^{1}$ School of Medicine, Hirosaki University
5 Zaifu-cho Hirosaki city, Aomori Prefecture, 036-8562, Japan\\
$^{2}$ Research Center for Advanced Science and Technology, The University of Tokyo,\\
4-6-1 Komaba, Meguro-ku, Tokyo 153-8904, Japan\\
$^{3}$ Department of Aeronautics and Astronautics, School of Engineering, The University of Tokyo,\\
7-3-1 Hongo, Bunkyo-ku, Tokyo 113-8656, Japan
}%

\date{\today}% It is always \today, today,

\begin{abstract}
\noindent
We develop a modified version of the totally asymmetric simple exclusion process (TASEP) and use it to reproduce flow on an escalator with two distinct lanes of pedestrian traffic. The model is used to compare strategies with two standing lanes and a standing lane with a walking lane, using theoretical analysis and numerical simulations. The results show that two standing lanes are better for smoother overall transportation, while a mixture of standing and walking is advantageous only in limited cases that have a small number of pedestrians. In contrast, with many pedestrians, the individual travel time of the first several entering particles is always shorter with distinct standing and walking lanes than it is with two standing lanes.
\end{abstract}

\maketitle

%\tableofcontents

\section{INTRODUCTION}
An escalator is an essential system for pedestrian transportation in many public facilities, such as train stations, shopping malls, and airports, that enable people to efficiently move. The capacity of an escalator has been investigated throughly~\cite{elevator1971escalators, strakosch1998vertical}. Such studies mainly focus on the maximum capacity of the system. 

When considering the operation of an escalator after it is installed, individual pedestrian behaviors must also be understood. In many countries, for example, etiquette dictates that people should stand on one side and walk on the other side of an escalator~\cite{mason2011walk,toda2018simulation,japan,bbc,age}. Somewhat counter-intuitively, however, some practitioners have recently started to encourage people to stand on both sides for smoother transportation and safety~\cite{kukadia2016pilot,JR,nytimes}. 

Research about escalators in the fields of transportation engineering and nonequilibrium statistical mechanics has only recently begun~\cite{kauffmann2011traffic,li2015simulation,al2016bottleneck,cheung1998pedestrian,ji2013study,kauffmann2013modeling,yue2018cellular} . Individual behavior has also been considered. Refs.~\cite{kauffmann2011traffic,li2015simulation,al2016bottleneck} investigate the flow of traffic mainly from numerical simulations, while Refs.~\cite{cheung1998pedestrian,ji2013study} study pedestrian choices between escalators and stairways. Especially, in Refs.~\cite{kauffmann2013modeling,yue2018cellular}, they investigate the escalator etiquette (standing on one side and walking on the other side) and conclude that walking is not beneficial in some cases. However, their results are basically obtained using only numerical simulations with some specific parameters. We consider it to be essential to anew investigate which escalator strategy is suitable for various situations using numerical simulations with extensive parameters and theoretical analyses.

In the present study, we analyze a novel escalator model that reproduces individual pedestrians' behaviors on both macro- (total transportation time or flow) and micro-scales (individual transportation time). Theoretical analysis and numerical simulations are used. To present an escalator, we construct a two-lane model with a modified totally asymmetric simple exclusion process (TASEP), which is a stochastic process on a one-dimensional lattice in which particles are allowed to hop in one direction (left to right in the present study). In the field of nonequilibrium statistical mechanics, researchers have applied TASEP extensively to various themes such as molecular-motor traffic~\cite{parmeggiani2003phase, chowdhury2005physics, chou2011non, appert2015intracellular}, vehicular traffic~\cite{RevModPhys.73.1067, PhysRevE.91.062818,PhysRevE.89.042813,PhysRevE.91.062818,yamamoto2017velocity}, and exclusive queuing processes~\cite{yanagisawa2010excluded, arita2010exclusive, arita2015exclusive}, and the process is especially useful since it can be solved exactly~\cite{PhysRevE.59.4899, 1751-8121-40-46-R01, 0305-4470-26-7-011}. Our model differs from the original TASEP with open boundaries in two respects.

First, the updating rules for particles are different. Specifically, in addition to the original hopping probability, particles can deterministically hop one site forward even when the right-neighboring site is occupied, which is an important feature of an escalator. 

Second, our model consists of two lanes that can have different hopping probabilities. The multi-lane TASEP has itself been investigated vigorously ~\cite{evans2011phase,wang2014phase,hilhorst2012multi,reichenbach2007traffic}; however, most of them assume the same hopping probability in all lanes. Fundamental behaviors of this model like the phase transitions are investigated in Refs.~\cite{evans2011phase,wang2014phase}. Meanwhile, Refs.~\cite{hilhorst2012multi,reichenbach2007traffic} analyze actual traffic flows using the multi-lane TASEP. 
We emphasize that most studies of the multi-lane model allow particles to switch lanes, while our model prohibits this behavior.

In the present study, we investigate three escalator strategies; (i) Strategy SS: two standing lanes, (ii) Strategy SW: one standing lane with one walking lane, and (iii) Strategy WW: two walking lanes. We note that Strategy SS and SW are our focus, because they are more common.

Using both theoretical analysis and numerical simulations with this model, we find that Strategy SS is generally more advantageous in terms of reducing the total (macro-scale) transportation time, especially with a relatively large number of particles, which model pedestrians. Conversely, in terms of reducing the individual (micro-scale) transportation time, Strategy SW offers better results for the first-entering particles, which tend to prefer walking, on the escalator.  

The rest of this paper is organized as follows. Section \ref{sec:model} defines the modified two-lane TASEP. Then Sec. \ref{sec:onelane} discusses the behavior of the basic one-lane model with our modified update rules. In Sec. \ref{sec:twolaneT}, we examine the total transportation time given with our modified two-lane model. Then, we proceed to discussion of individual transportation times in Sec. \ref{sec:twolanet}. Finally, Sec. \ref{sec:conclusion} gives concluding remarks. 

\section{MODEL DESCRIPTION}
\label{sec:model}
\subsection{Original TASEP with open-boundary conditions}
\begin{figure}[htbp]
\begin{center}
\includegraphics[width=8.5cm,clip]{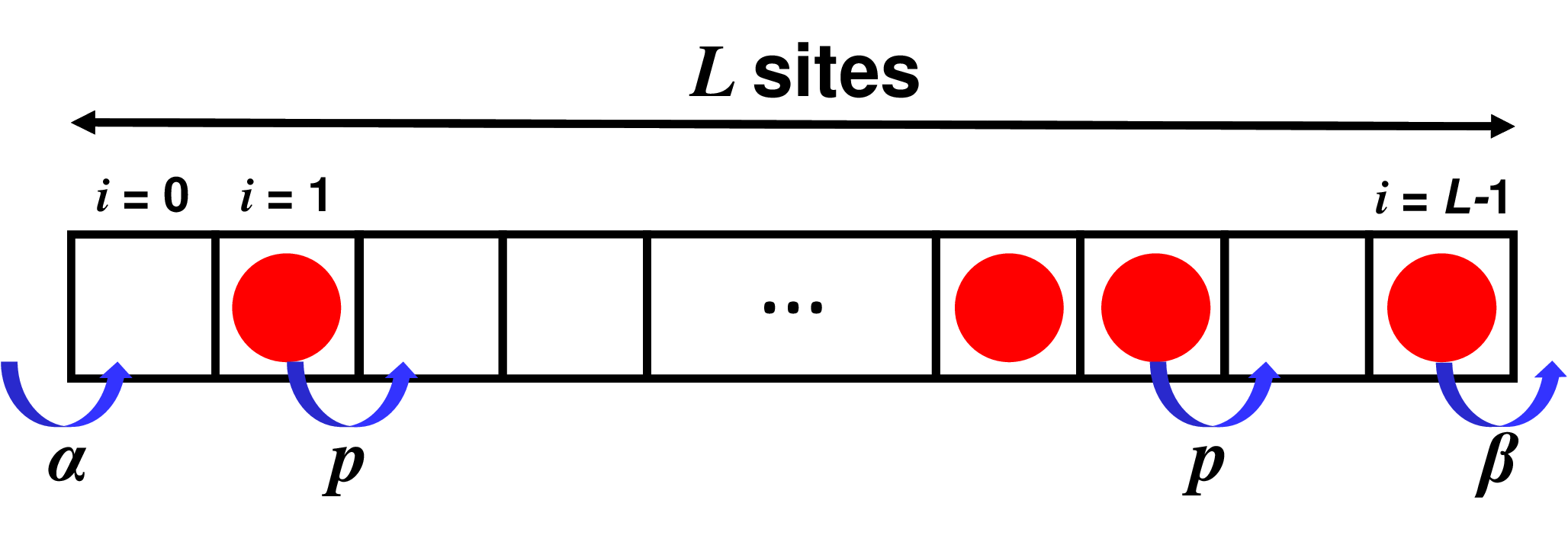}
\caption{(Color Online) Schematic illustration of the original TASEP with open-boundary conditions.}
\label{fig:original_TASEP}
\end{center}
\end{figure}

The original TASEP with open-boundary conditions is defined as a one-dimensional lattice (lane) of $L$ sites, which are labeled from left to right as $i=0,1,......,L-1$, as illustrated in Fig. \ref{fig:original_TASEP}. Each site can be either empty or occupied by only one particle. When the $i$th site at time $t$ is occupied by a particle, its state is represented as $s_i(t)=1$; otherwise, its state is $s_i(t)=0$. 

Discrete time and parallel updating are adopted in the present study. During parallel updating, the states of all the particles on the lattice are determined simultaneously in the next time step. Particles enter the lattice from the left boundary with input probability $\alpha$, and they leave the lattice from the right boundary with output probability $\beta$. In the bulk, particles whose right-neighboring sites are empty can hop to the rightward site with probability $p$; otherwise they remain at their present site. The modified TASEP differs from the original process in the following two respects.

\subsection{Modification 1: Updating rules}
\label{sec:difference1}
Our modified updating rules have two-step structure. First, particles hop deterministically one site forward regardless of the right-neighboring site, which reproduces an important feature of an escalator. 

In addition, particles may hop one more site forward with hopping probability $p$, which is fixed throughout each lattice in our model, if the right-neighboring site is empty. This possibility represents walking on an escalator. 

So, particles can hop one or two sites for each time step. We note that in our model particles have two opportunities for hopping and can hop one or two sites in each time step, unlike the Nagel-Schreckenberg model for vehicular traffic~\cite{nagel1992cellular}, in which particles hop equal to or more than zero site only once in each time step. Table \ref{fig:updating rules} summarizes the modified updating rules in comparison with the original updating rules.

\renewcommand{\figurename}{TABLE}
\renewcommand{\thefigure}{\Roman{figure}}
\setcounter{figure}{0}
\begin{figure}[htbp]
\begin{center}
\caption{(Color Online) Updating rules for the red particles in the bulk of our model for comparison with the original TASEP. The notation `Prob.' represents the probability of each configuration at time ($t+1$). We note that blue particles are not depicted at time ($t+1$).}
\includegraphics[width=8.5cm,clip]{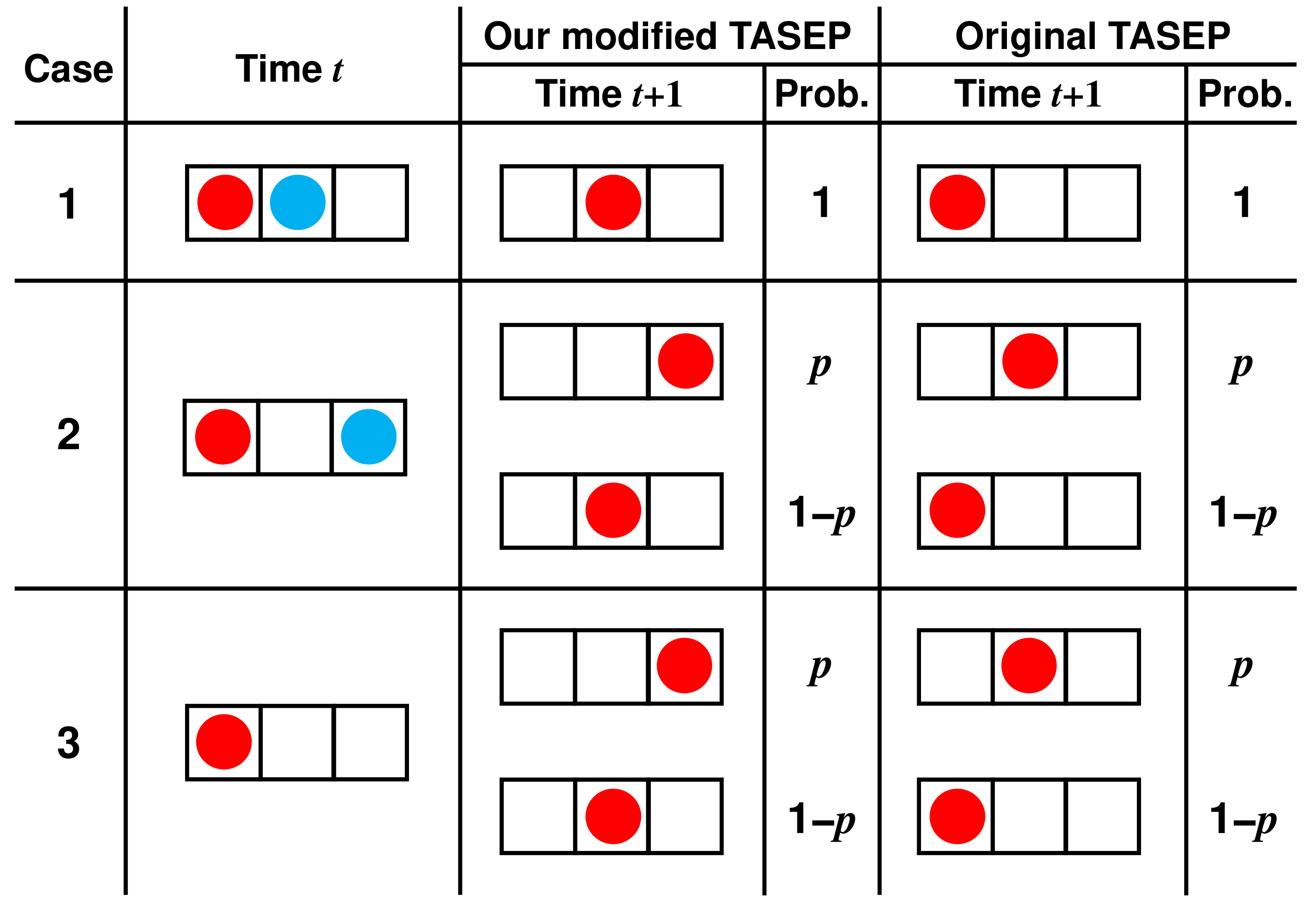}
\label{fig:updating rules}
\end{center}
\end{figure}

\begin{figure}[htbp]
\begin{center}
\caption{(Color Online) Updating rules for the red particles around the right boundary.}
\includegraphics[width=8.5cm,clip]{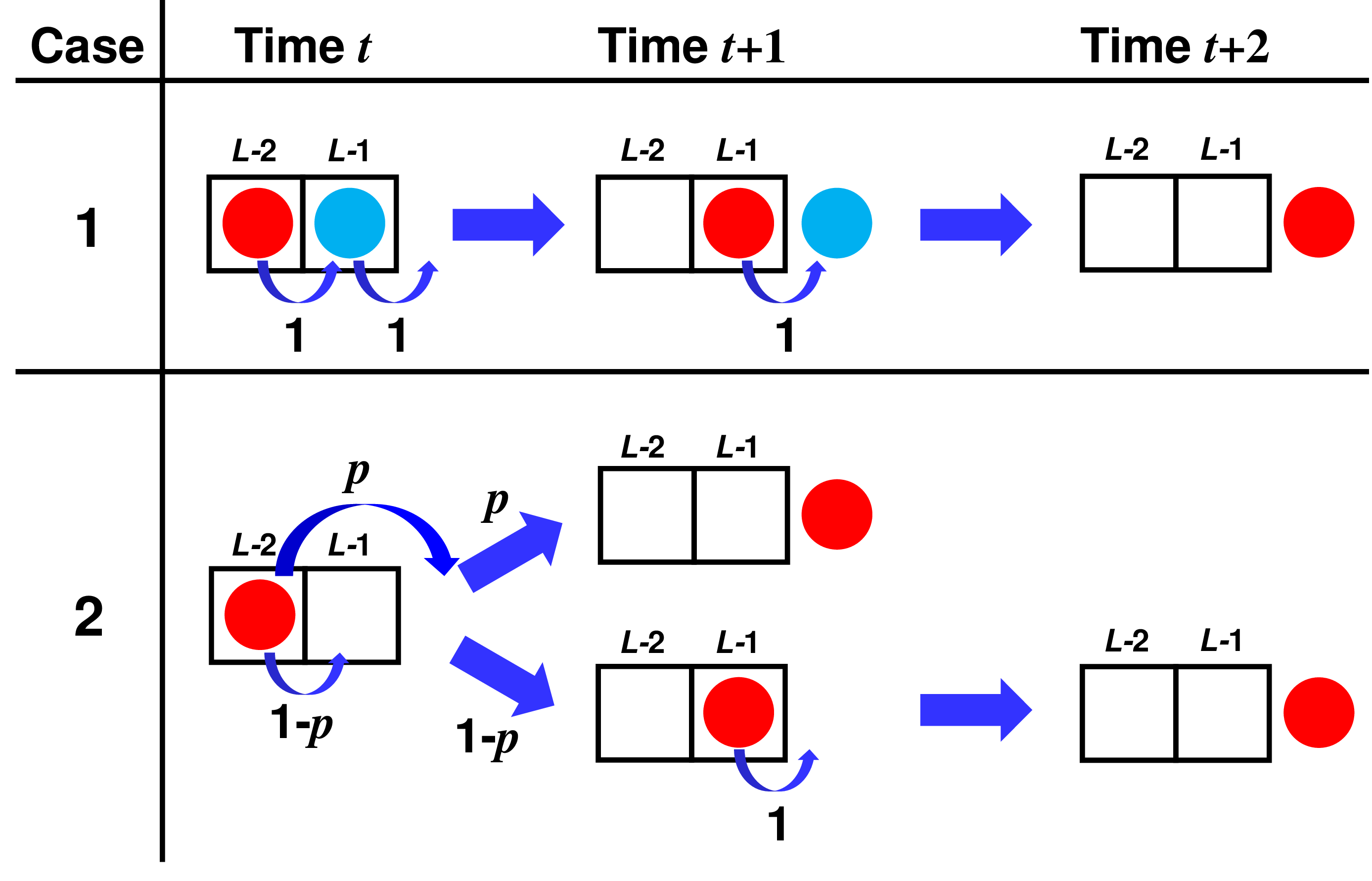}
\label{fig:exit}
\end{center}
\end{figure}

At the right boundary, a particle can leave the system from the ($L-1$)th site or from the ($L-2$)th site, unlike in the original TASEP, in which only a particle occupying the ($L-1$)th site leaves the system with output probability $\beta$. In the present model, a particle occupying the ($L-2$)th site at time $t$ leaves the system with probability $p$ and hops to the ($L-1$)th site with $1-p$ at time ($t+1$) if $s_{L-1}(t)=0$; otherwise it hops to the ($L-1$)th site. On the other hand, a particle occupying the ($L-1$)th site at time $t$ must leave the system. Table \ref{fig:exit} extracts and summarizes the updating rules around the right boundary.

\renewcommand{\figurename}{FIG.}
\renewcommand{\thefigure}{\arabic{figure}}

\subsection{Modification 2: Two lanes with two types of particles}
\label{sec:difference2}
Second, our modified model consists of two lanes. The state of the $i$th site of the lane 1 and 2 at time $t$ are represented as $s_i^1(t)$ and $s_i^2(t)$, respectively. Each lane can act as a standing lane ($p=0$) or a walking lane ($0<p\leq 1$). Changing lanes is prohibited. 

\setcounter{figure}{1}
\begin{figure}[htbp]
\begin{center}
\includegraphics[width=8.5cm,clip]{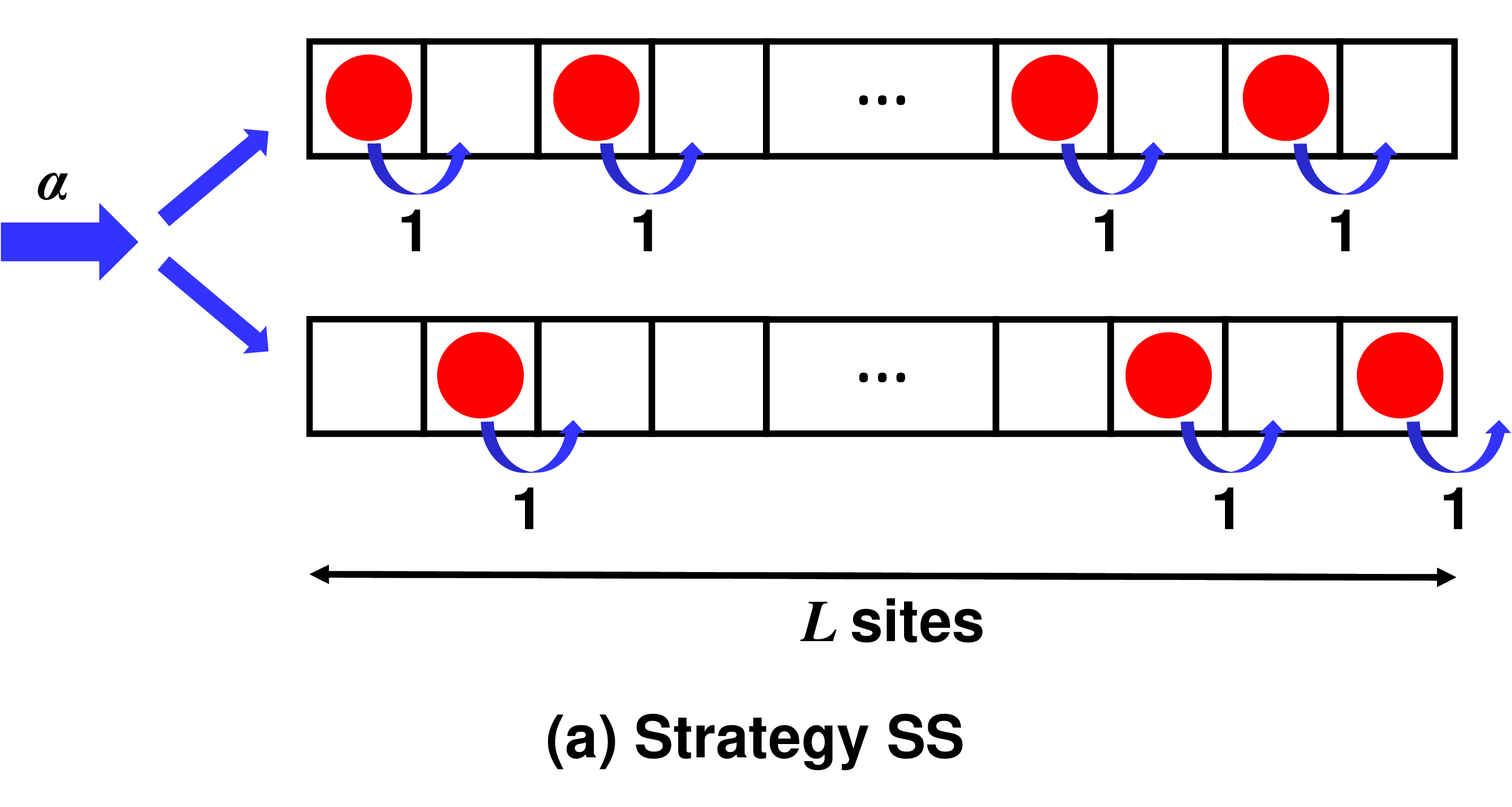}\\
\includegraphics[width=8.5cm,clip]{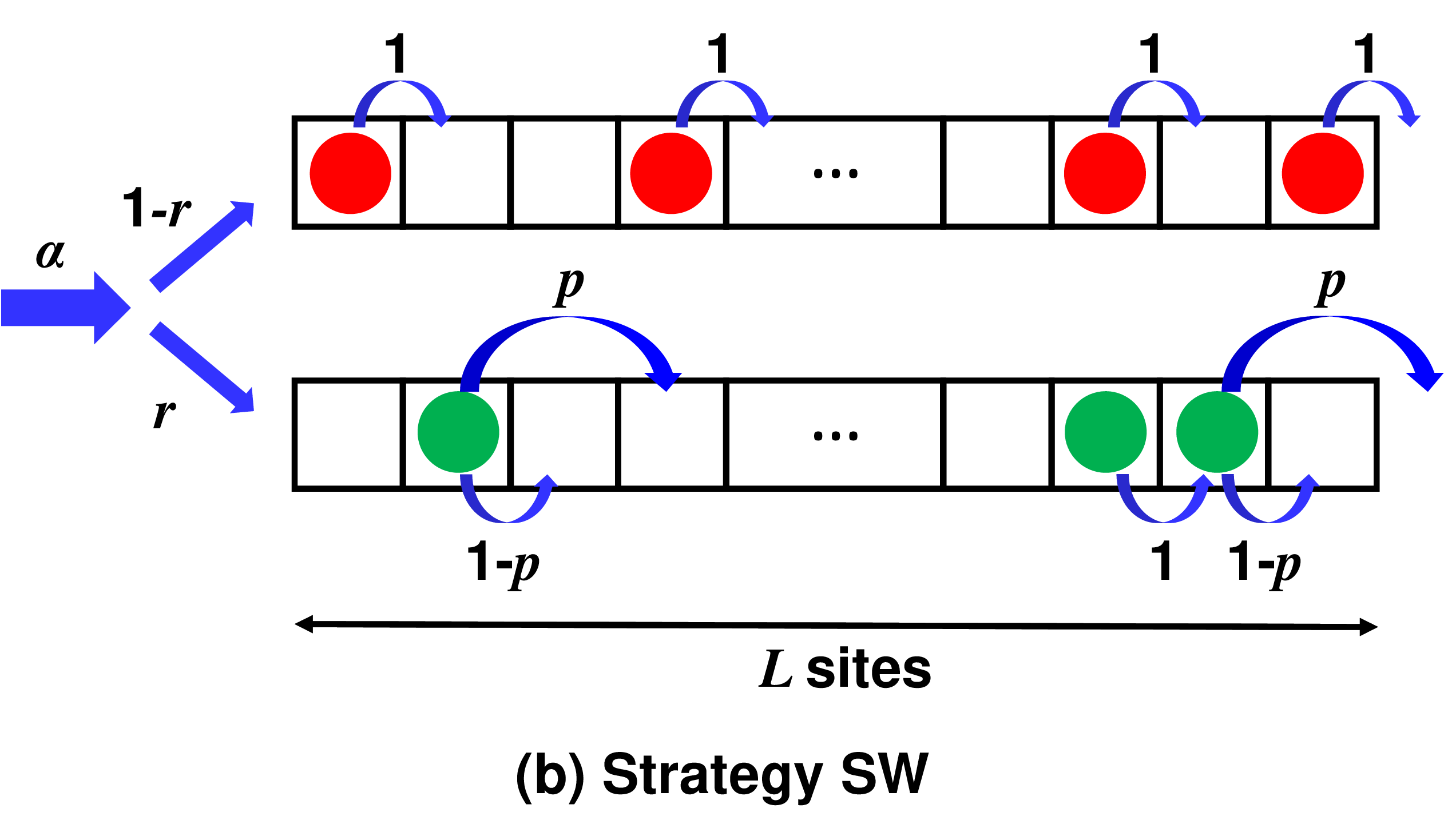}\\
\includegraphics[width=8.5cm,clip]{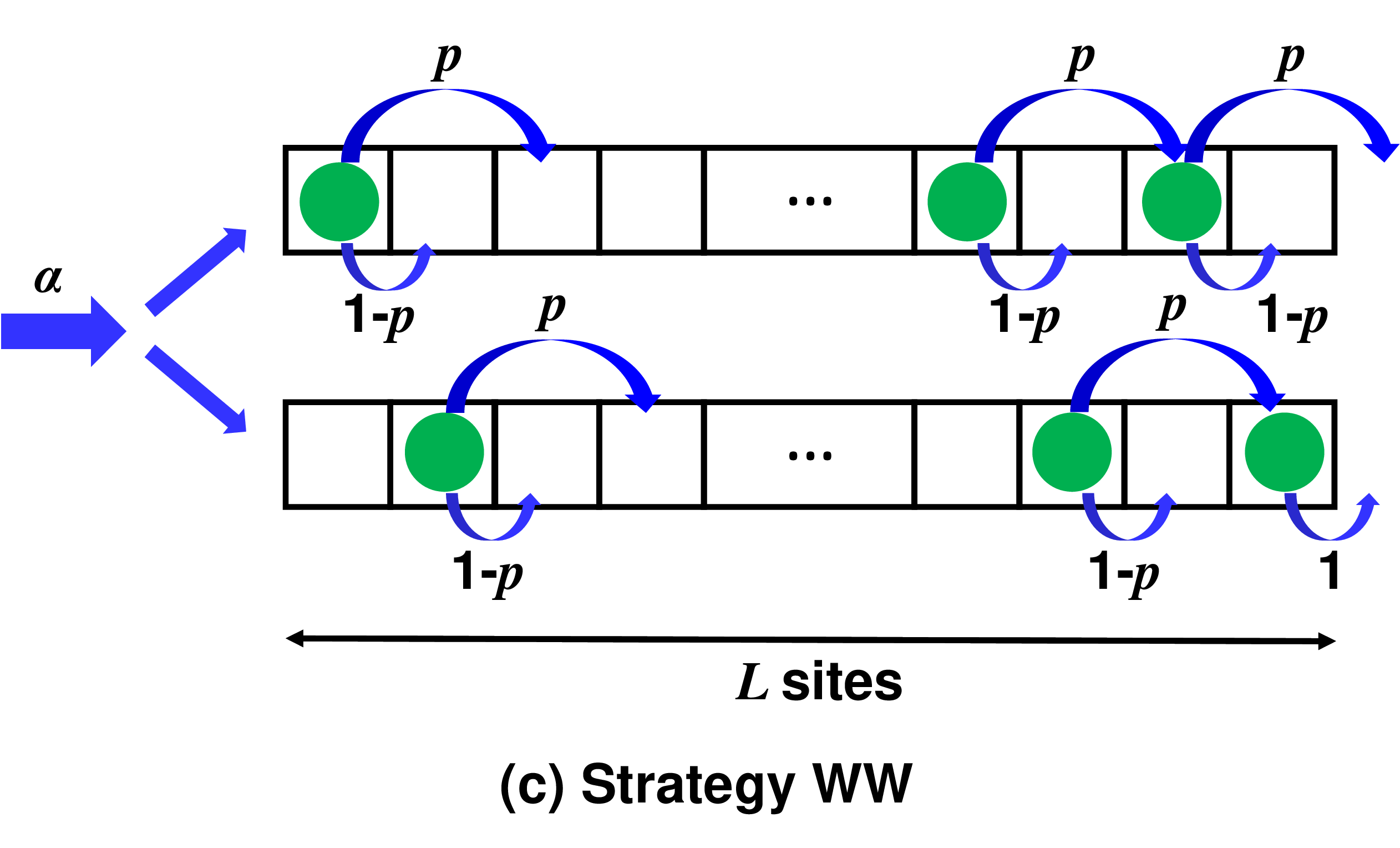}
\caption{(Color Online) Schematic illustration of the modified TASEP with (a) Strategy SS: two standing lanes, (b) Strategy SW: one standing lane (upper) and one walking lane (lower), and (c) Strategy WW: two walking lanes. We note that red (green) particles prefer standing (walking).}
\label{fig:lattice}
\end{center}
\end{figure}

Now, we consider three strategies: (1) two standing lanes, which corresponds to Strategy SS (see Fig. \ref{fig:lattice} (a)), (2) one standing lane with one walking lane, corresponding to Strategy SW (see Fig. \ref{fig:lattice} (b)), and (3) two walking lanes, corresponding to Strategy WW (see Fig. \ref{fig:lattice} (c)). Strategy WW is generally not adopted in real situations; however, it is investigated here for the sake of comparison.

With Strategy SS and WW, particles can enter either of the two lanes if the first site is empty. Specifically, a particle enters lane 1 (lane 2) if $(s_i^1(t),s_i^2(t))=(0,1)$ (if $(s_i^1(t),s_i^2(t))=(1,0)$), while it must select enter either of two lanes if $(s_i^1(t),s_i^2(t))=(0,0)$ indiscriminately, i.e., with probability 1/2.

We emphasize that particles can always enter the system with Strategy SS and WW. This is because all the particles in the system will hop one or two sites forward every time step, so at least one of the two left boundaries will always be vacant; that is, $(s_1^1(t),s_1^2(t))=(0,1)$, $(1,0)$, or $(0,0)$.

With Strategy SW, however, two types of particles are possible; waking-preference particles with probability $r$ and standing-preference particles with probability $1-r$. Standing (walking) particles can enter the standing (walking) lane if the leftmost site of the corresponding lane is empty; otherwise they cannot. Unlike Strategy SS and WW, particles frequently cannot enter the system in this case because of the preference. We note that in the simulations below the preference of a particle is determined just before it enters the system; this preference is reset if the particle cannot enter and is redetermined at the next chance of entering.

\section{One-lane model \\ with the modified updating rules}
\label{sec:onelane}
In this section, we briefly discuss the steady-state flow of the basic one-lane model with the modified updating rules. 

\begin{figure}[htbp]
\begin{center}
\includegraphics[width=8.5cm,clip]{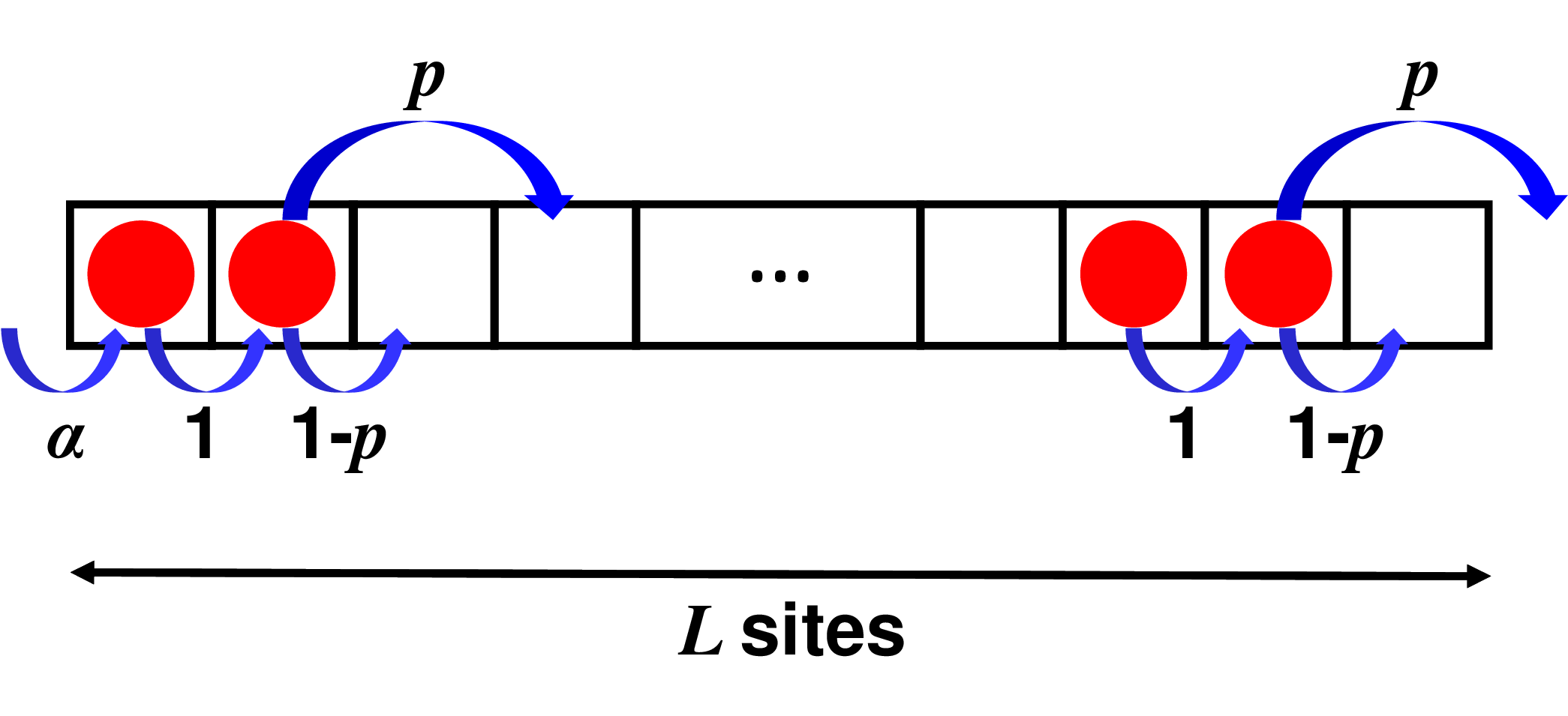}
\caption{(Color Online) Schematic illustration of the one-lane model with our modified updating rules.}
\label{fig:onelattice}
\end{center}
\end{figure}

The basic one-lane model with our modified updating rules, as illustrated in Fig. \ref{fig:onelattice}. In this system, particles must hop one or two sites forward. Therefore, the steady-state flow is clearly equal to or more than that of the original TASEP with hopping probability $p=1$. 

The original TASEP with $p=1$ for various ($\alpha,\beta$) exhibits only two phases; the low-density (LD) phase, in which the system is governed by the left boundary, and the high-density (HD) phase, in which the system is governed by the right boundary, and the maximal current (MC) phase, in which the system is governed by the bulk, never occurs. Therefore, the one-lane model is always governed by the left boundary, since a queue can never form near the right boundary. This fact, counter-intuitively, indicates that the flow is determined regardless of $p$, so the flow is always constant. This can be explained as follows. 

\begin{figure}[htbp]
\begin{center}
\includegraphics[width=8.5cm,clip]{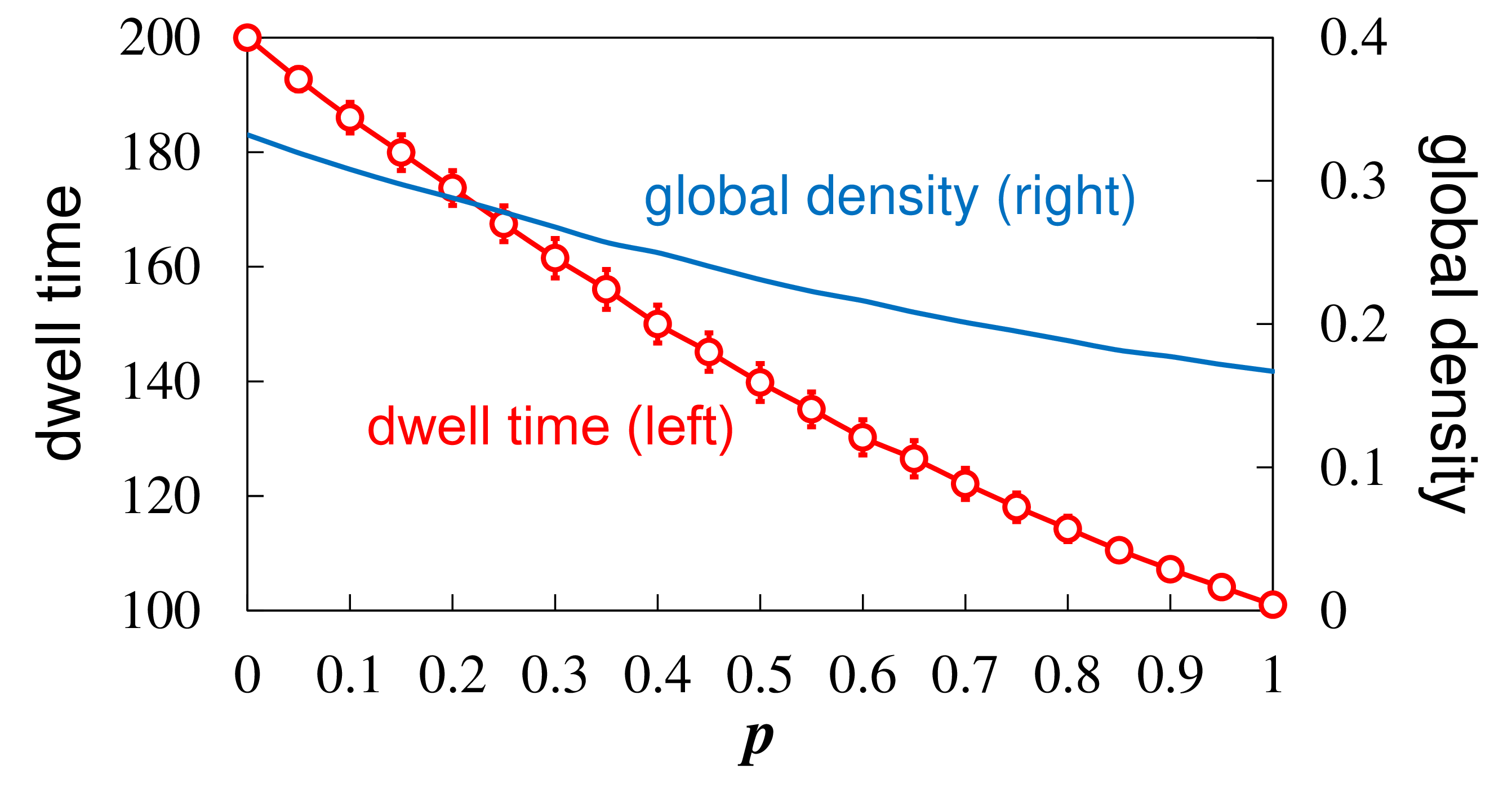}
\caption{(Color Online) Simulation values of dwell time (red, left axis) with the sample standard deviation and global density (blue, right axis) as functions of $p$. The dwell time is calculated by averaging the times of 1000 particles starting from the $(10^4+1)$th particle, and the global density calculated by averaging over $10^5$ time steps after evolving the system for $10^4$ time steps.}
\label{fig:dwelltime}
\end{center}
\end{figure}

First, walking clearly reduces the dwell time of each particle in the system, which is defined as the time gap between the time when the particle enters the system and the time when it leaves. The dwell time decreases as $p$ increases. In addition, the global density of the lane, which is defined as the average number of occupied cells over the space $[0,L-1]$ in one time step, decreases as $p$ increases, due to the longer gap between particles. Figure \ref{fig:dwelltime} shows the average dwell time and the global density as functions of $p$, and reproduce the behavior discussed above. Since (i) the average velocity of particles is proportional to the dwell time, and (ii) the flow is represented as a multiplication of the average velocity and the global density, the flow remains constant.

The steady-state flow $Q_1$ of the one-lane model with our updating rules satisfies 
\begin{equation}
Q_1=\alpha(1-Q_1),
\end{equation}
resulting in
\begin{equation}
Q_1=\frac{\alpha}{1+\alpha}.
\label{eq:flow}
\end{equation}
Eq. (\ref{eq:flow}) is equal to that of the original TASEP in the LD phase with parallel updating, indicating that the one-lane system always exhibits the LD phase. 

\begin{figure}[htbp]
\begin{center}
\includegraphics[width=8.5cm,clip]{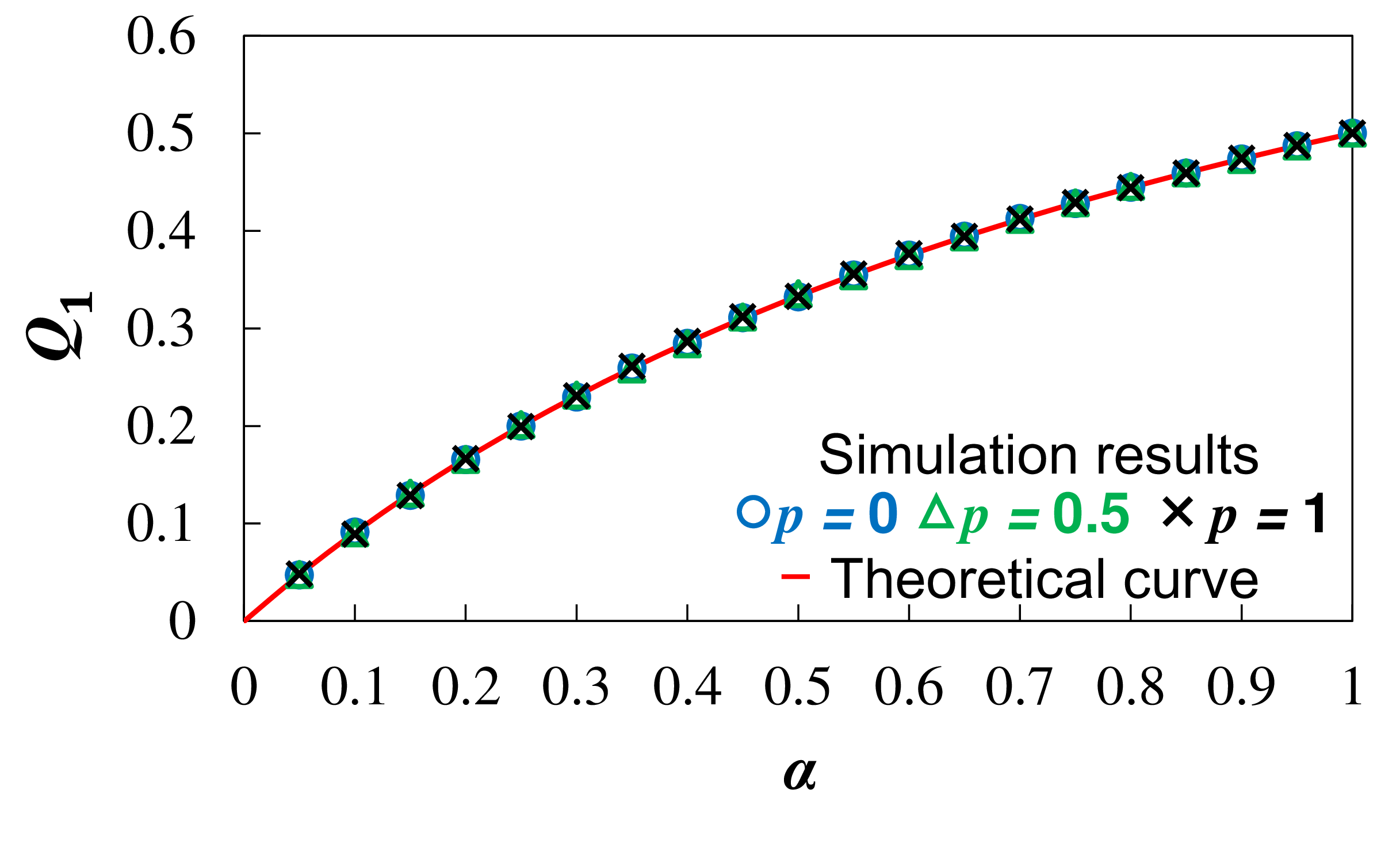}
\caption{(Color Online) Simulation (dots) and theoretical (curve) values of $Q_1$ as functions of $\alpha$ for various $p\in\{0  {\rm (blue \ circles)},0.5  {\rm (green \  triangles)},1  {\rm (black \ crosses)}\}$.}
\label{fig:Qone}
\end{center}
\end{figure}

Figure \ref{fig:Qone} compares the simulation (dots) and theoretical (curve) values of $Q$ as functions of $\alpha$ for various $p\in\{0,0.5,1\}$. In the simulations, we set $L=200$ and the flow is obtained by averaging over $10^5$ time steps after evolving the system for $10^4$ time steps (and similarly hereafter).  

In Fig. \ref{fig:Qone}, the simulation values show very good agreement with the theoretical curve, and at the same time, the simulations also confirm that $Q_1$ is independent on $p$.

\section{Total (macro-scale) \\ transportation time \\ with the two-lane model}
\label{sec:twolaneT}
In this section, we investigate the total (macro-scale) transportation time $T$ of $N$ particles with the two-lane model. Specifically, $T$ is defined as the time gap between the start of the simulation and the time at which the final leaving particle leaves the system. The total transportation times with Strategy SS, SW, and WW are represented as $T_{\rm SS}$, $T_{\rm SW}$, and $T_{\rm WW}$, respectively.

\subsection{Steady-state flow $Q_{\rm SS}$, $Q_{\rm SW}$, and $Q_{\rm WW}$}
Before examining $T$, we briefly discuss the steady-state flow in the two-lane model. 

Since particles can always enter the system with Strategy SS and WW (see Subsec. \ref{sec:difference2}), the following relation holds  
\begin{equation}
Q_{\rm SS}=Q_{\rm WW}=\alpha,
\label{eq:QSS}
\end{equation}
where $Q_{\rm SS}$ and $Q_{\rm WW}$ are defined as the steady-state flow of the two-lane model with Strategy SS and WW, respectively. We emphasize that counter-intuitively, $Q_{\rm WW}=Q_{\rm SS}$ because of the independence of the flow from $p$ (see the previous section), so walking can have no effect on increasing the steady-state flow.

Second, with Strategy SW, particles prefer to enter a standing lane or a walking lane with some probability. Given that a particle prefers standing (walking) with probability $1-r$ ($r$), the input probability of the standing (walking) lane reduces to $(1-r)\alpha$ ($r\alpha$). Therefore, the steady-state flows of the standing lane $Q_{\rm S}$ and the walking lane $Q_{\rm W}$ in the two-lane model are given as 

\begin{equation}
Q_{\rm S}=\frac{(1-r)\alpha}{1+(1-r)\alpha},
\label{eq:QS}
\end{equation}
and
\begin{equation}
Q_{\rm W}=\frac{r\alpha}{1+r\alpha},
\label{eq:QW}
\end{equation}
which are derived from Eq. (\ref{eq:flow}) by replacing $\alpha$ with $(1-r)\alpha$ and $r\alpha$, respectively. 

Consequently, the steady-state flow of the two-lane model with Strategy SW, $Q_{\rm SW}$, can be calculated as 

\begin{equation}
Q_{\rm SW}=Q_{\rm S}+Q_{\rm W}=\frac{(1-r)\alpha}{1+(1-r)\alpha}+\frac{r\alpha}{1+r\alpha},
\label{eq:QSW}
\end{equation}
which takes its maximum value when $r=0.5$. The detailed properties of the function of Eq. (\ref{eq:QSW}) are discussed in Appendix \ref{sec:QSW}. 

Figure \ref{fig:Qcompare} compares the simulation (dots) and theoretical (curves) values of (a) $Q_{\rm SS}$, $Q_{\rm SW}$ for $p\in\{0.5, 1\}$, and $Q_{\rm WW}$ for $p\in\{0.5, 1\}$ as functions of $\alpha$, and (b) $Q_{\rm SW}$ for various $\alpha\in\{0.2, 0.6, 1\}$ and $p\in\{0.5, 1\}$ as functions of $r$. The simulation values again show very good agreement with the theoretical curves. 

\begin{figure}[htbp]
\begin{center}
\includegraphics[width=8.2cm,clip]{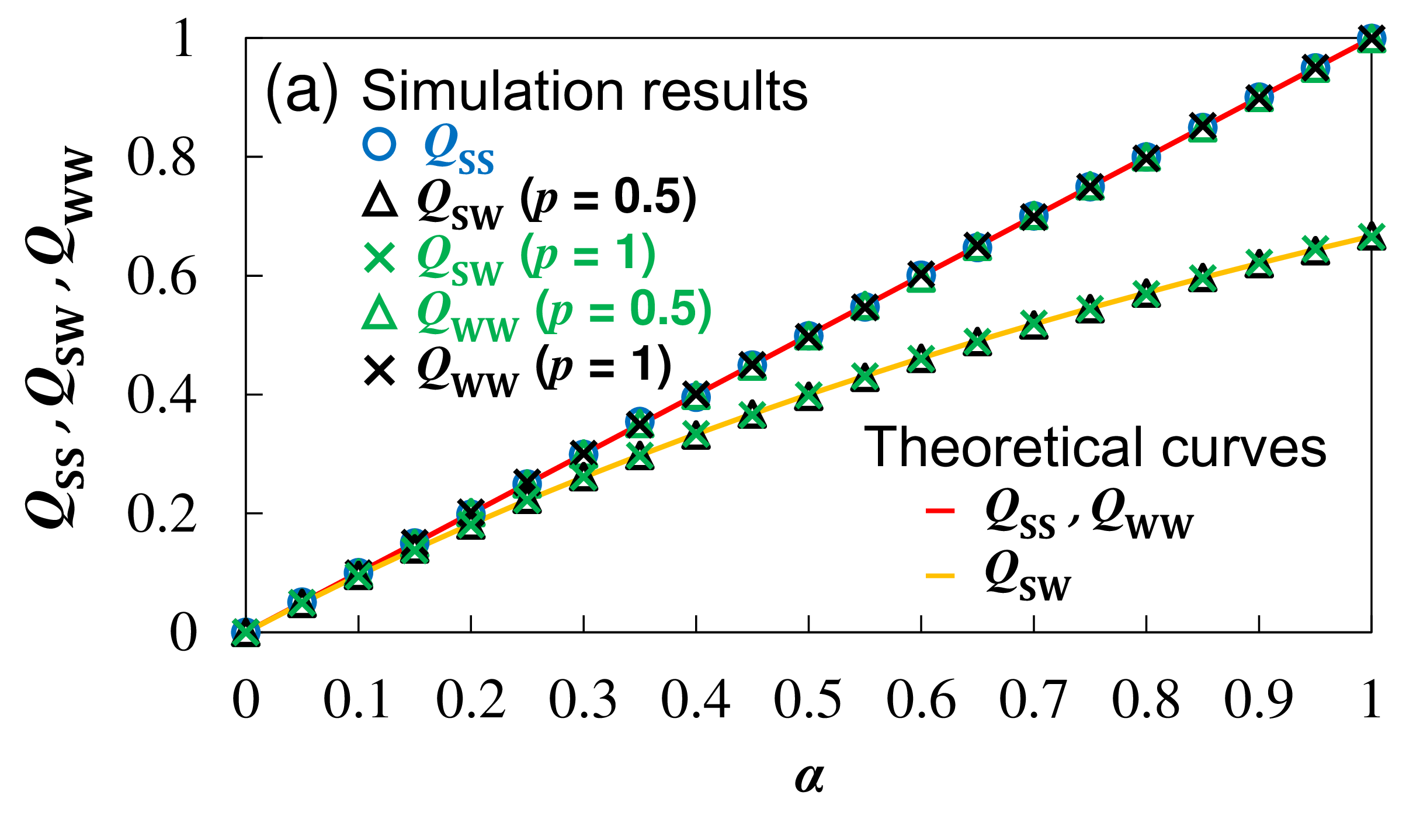}\\
\includegraphics[width=8.2cm,clip]{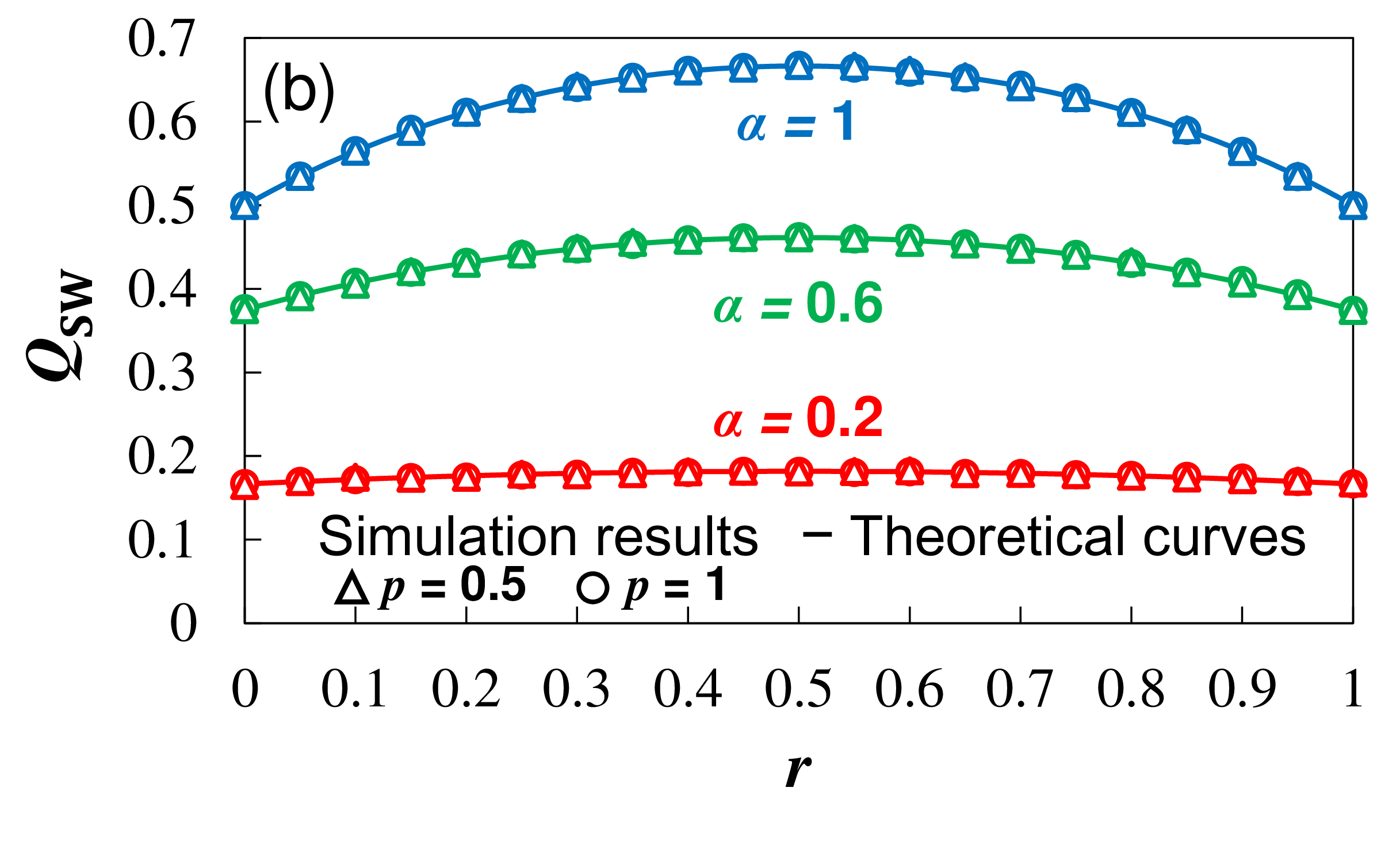}
\caption{(Color online) Simulation (dots) and theoretical (red/orange \ curves) values of (a) $Q_{\rm SS}$ (blue \ circles), $Q_{\rm SW}$ for $p\in\{0.5  {\rm (black \ triangles)}, 1  {\rm (green \ crosses)}\}$, and $Q_{\rm WW}$ for $p\in\{0.5  {\rm (green \  triangles)}, 1  {\rm (black \ crosses)}\}$ as functions of $\alpha$, and (b) $Q_{\rm SW}$ for various $\alpha\in\{0.2  {\rm (red)}, 0.6  {\rm (green)}, 1  {\rm (blue)}\}$ and $p\in\{0.5  {\rm (triangles)}, 1  {\rm (circles)}\}$ as functions of $r$.}
\label{fig:Qcompare}
\end{center}
\end{figure}

\renewcommand{\figurename}{TABLE}
\renewcommand{\thefigure}{\Roman{figure}}
\setcounter{figure}{2}
\begin{figure}[htbp]
\begin{center}
\caption{(Color Online) Comparison of the left boundary between Strategy SS and SW, with constant $\alpha=1$. Two red particles can enter either of two standing lanes with Strategy SS. Two green particles, which prefer walking in this figure, attempt to enter the walking lane with Strategy SW. We note that the upper (lower) lane is a standing (walking) lane for Strategy SW.}
\includegraphics[width=8.1cm,clip]{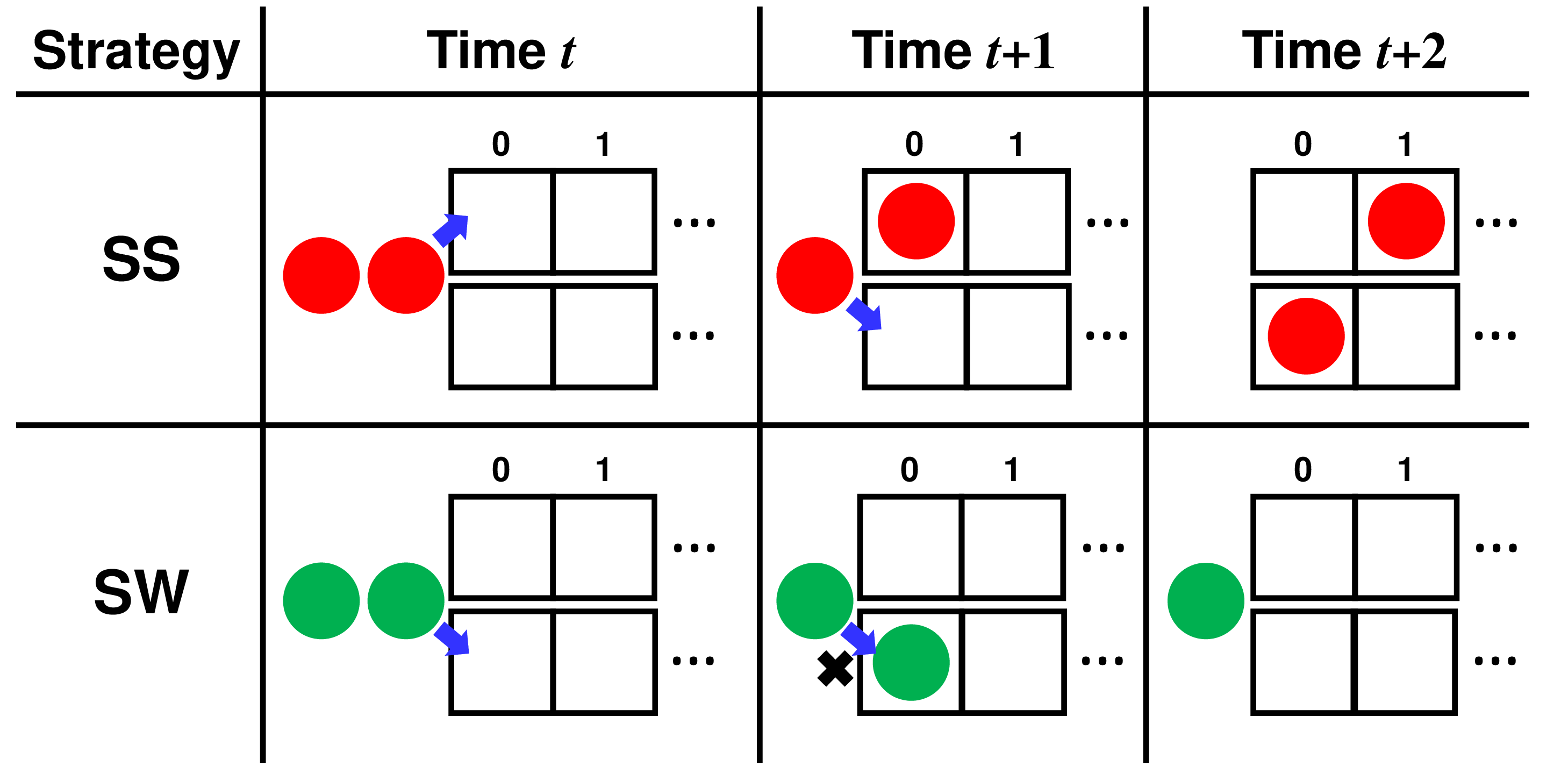}
\label{fig:1vs2}
\end{center}
\end{figure}
\renewcommand{\figurename}{FIG.}
\renewcommand{\thefigure}{\arabic{figure}}
\setcounter{figure}{6}

From Eqs. (\ref{eq:QSS}) and (\ref{eq:QSW}), we immediately obtain
\begin{equation}
Q_{\rm SW}<r\alpha+(1-r)\alpha=\alpha=Q_{\rm SS}(=Q_{\rm WW}),
\label{eq:QSWQSS}
\end{equation}
indicating that Strategy SS (WW) is always advantageous in terms of steady-state flow. 

The absolute advantage of Strategy SS (WW) over Strategy SW is explained with the behavior at the entrances of the lanes. Table \ref{fig:1vs2} summarizes the model behavior at the left boundary, extracting the first and second site with $\alpha=1$.

With Strategy SS (WW), since consecutive pairs of particles mutually enter either lane, two particles can enter the system in two time steps, as shown in the upper panel of Tab. \ref{fig:1vs2}. 

On the other hand, with Strategy SW they cannot enter the system with two time steps but with three time steps, if both of two consecutive particles' preference are the same, as shown in the lower panel of Tab. \ref{fig:1vs2}. The two particles can enter the system in two time steps if two consecutive particles separately prefer walking and standing. Therefore, $Q_{\rm SW}$ is maximized if $r=0.5$, which clearly maximizes the probability that the preferences of the two consecutive particles differ. 

Considering that the system is governed by the left boundary, these facts finally lead to $Q_{\rm SS}>Q_{\rm SW}$, explaining the absolute advantage of Strategy SS (WW) over Strategy SW.

\subsection{Total (macro-scale) transportation time $T$}
\subsubsection{Approximate theoretical analyses of $T$}
In this subsection, we theoretically calculate the approximations of $T_{\rm SS}$ (total transportation time with Strategy SS), $T_{\rm SW}$ (total transportation time with Strategy SW), and $T_{\rm WW}$ (total transportation time with Strategy WW) and consider the relations among them.

Considering that the steady-state flow $Q$ expresses the average number of particles that pass a certain point (the left boundary) each time step, the average required time steps $T^{\rm s}(N)$ for the $N$th particle to enter the system from the time at which the first-entering particle enters in steady-state flow can be calculated as 
\begin{equation}
T^{\rm s}(N)\approx\frac{N-1}{Q}.
\label{eq:Ts}
\end{equation} 

On the other hand, the required time steps $T^{\rm f}(p)$ for the final-leaving particle of both lanes, which is not always identical to the final-entering particle, to reach the right boundary can be represented as 
\begin{eqnarray}
\left\{ \begin{array}{ll}
T^{\rm f}=T_{\rm S}^{\rm f}=\displaystyle\frac{L}{1} & {\rm for \ a \ standing \ lane,}  \\
T^{\rm f}(p)=T_{\rm W}^{\rm f}(p)\gtrapprox\displaystyle\frac{L}{1+p} & {\rm for \ a \ walking \ lane,}
\end{array} \right.
\label{eq:Tf}
\end{eqnarray}
where $T_{\rm S}^{\rm f}$ ($T_{\rm W}^{\rm f}$) is a (an) definitive (approximate) value, and when calculating $T_{\rm W}^{\rm f}$ the possibility that a walking particle is blocked by the particle ahead of it is ignored, resulting in a slight underestimation of $T_{\rm W}^{\rm f}$ (see also Sec. \ref{sec:validityT}).

Assuming that (i) $1/\alpha$ time steps are needed on average for the first particle to enter the system, and that (ii) particles enter both lanes during the steady-state flow after the first particle enters the system, $T_{\rm SS}(N,\alpha)$ and $T_{\rm WW}(N,\alpha,p)$--- their arguments can be abbreviated unless otherwise specified (similarly for other variables)---can be written as
\begin{equation}
T_{\rm SS}(N,\alpha)\approx \frac{1}{\alpha}+T_{\rm SS}^{\rm s}(N)+T_{\rm S}^{\rm f}=\frac{N}{\alpha}+\frac{L}{1}
\label{eq:TSS}
\end{equation}
and
\begin{equation}
T_{\rm WW}(N,\alpha,p)\approx \frac{1}{\alpha}+T_{\rm WW}^{\rm s}(N)+T_{\rm W}^{\rm f}=\frac{N}{\alpha}+\frac{L}{1+p},
\label{eq:TWW}
\end{equation}
from Eqs. (\ref{eq:QSS}), (\ref{eq:Ts}), and (\ref{eq:Tf}). We note that the first-in-first-out condition---i.e., the entering sequence is identical to the leaving sequence---must be satisfied when Strategy SS is modeled, whereas the relation is mostly satisfied when Strategy WW is modeled. 

From Eqs. (\ref{eq:TSS}) and (\ref{eq:TWW}), $T_{\rm SS}$ and $T_{\rm WW}$ differ only due to the difference in the second term, resulting in $T_{\rm SS}\approx T_{\rm WW}$ if $N$ is sufficiently large.

Under the same assumption, we then consider $T_{\rm SW}$. Unlike $T_{\rm SS}$ and $T_{\rm WW}$, the first-in-first-out condition generally does not hold since the hopping probabilities of the two lanes are different, leading to difficulty in calculating $T_{\rm SW}$. 

For approximate calculations of $T_{\rm SW}$, we need to consider the final-entering standing-preference particle and the final-entering walking-preference one. Let us define $N_0$ as a threshold number. Specifically, if a standing-preference particle is (not) included in the last $N_0$ entering particles, the final leaving particle is on average identical to the final-entering standing-preference (walking-preference) particle. We note that if $r=1$, $N_0$ is always $N_0=0$. 

\begin{figure}[htbp]
\begin{center}
\includegraphics[width=8.5cm,clip]{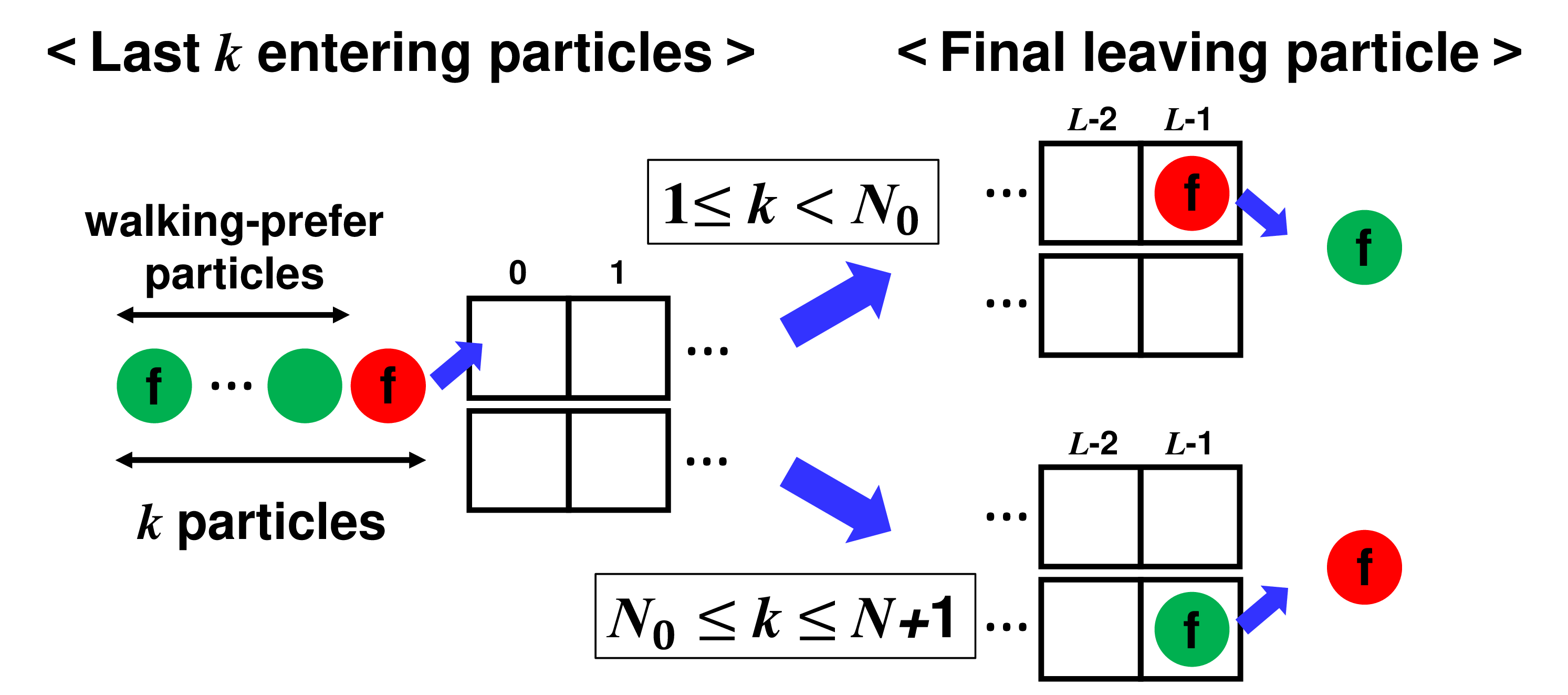}
\caption{(Color online) Schematic illustration for explaining $N_0$. The red (green) particles are standing (walking) particles, and the red (green) particle labeled `f' is the final-entering standing-preference (walking-preference) particle. If $1\leq k< N_0$, the final-leaving particle is, on average, identical to the final-entering standing-preference particle; otherwise the final-leaving particle is identical to the final-entering walking-preference particle.}
\label{fig:N0}
\end{center}
\end{figure}

Figure \ref{fig:N0} gives a schematic illustration of $N_0$. This figure focuses on the last $k$ entering particles, where the $k$th ($k=1,......,N+1$) entering particle from the final-entering particle is identical to the final-entering standing-preference particle. We note that $k=N+1$ represents that all $N$ particles prefer walking, which is defined for the sake of convenience.

\setcounter{equation}{15}
\begin{figure*}
\flushleft
\begin{eqnarray}
\begin{split}
T_{\rm SW}&\approx\displaystyle\sum_{k=1}^{N+1} P(k)t_{\rm SW}(k)
\\
&=\displaystyle\frac{1}{\alpha}+\displaystyle\sum_{k=1}^{N_1} \{P(k)\times(T_{\rm SW}^{\rm s}(N-k+1)+T_{\rm S}^{\rm f})\} + \left(1-\sum_{k=1}^{N_1} P(k)\right)\times(T_{\rm SW}^{\rm s}(N)+T_{\rm W}^{\rm f}) 
\\
&=\displaystyle\frac{1}{\alpha}+\displaystyle\frac{N-1}{Q_{\rm SW}}-\displaystyle\frac{\sum_{k=1}^{N_1}(1-r^{N_1-k})r^{k}}{Q_{\rm SW}}+\displaystyle\frac{(1-r^{N_1})L}{1}+\displaystyle\frac{r^{N_1}L}{1+p}
\\
&=\displaystyle\frac{1}{\alpha}+\displaystyle\frac{N-1}{\frac{r\alpha}{1+r\alpha}+\frac{(1-r)\alpha}{1+(1-r)\alpha}}-\displaystyle\frac{\sum_{k=1}^{N_1}(1-r^{N_1-k})r^{k}}{\frac{r\alpha}{1+r\alpha}+\frac{(1-r)\alpha}{1+(1-r)\alpha}}+\displaystyle\frac{(1-r^{N_1})L}{1}+\displaystyle\frac{r^{N_1}L}{1+p}
%&=\underbrace{\displaystyle\frac{1}{\alpha}+\displaystyle\frac{N-1}{\frac{r\alpha}{1+r\alpha}+\frac{(1-r)\alpha}{1+(1-r)\alpha}}-\displaystyle\frac{\sum_{k=1}^{N_1}(1-r^{N_1-k})r^{k}}{\frac{r\alpha}{1+r\alpha}+\frac{(1-r)\alpha}{1+(1-r)\alpha}}}_{\displaystyle{\rm first \ part}}\underbrace{\displaystyle+\frac{(1-r^{N_1})L}{1}+\displaystyle\frac{r^{N_1}L}{1+p}}_{\displaystyle{\rm second \ part}}
%&=\left\{ \begin{array}{ll}
%\underbrace{\displaystyle\frac{1}{\alpha}+\displaystyle\frac{N-1}{\frac{r\alpha}{1+r\alpha}+\frac{(1-r)\alpha}{1+(1-r)\alpha}}-\displaystyle\frac{\sum_{k=1}^{N_1}(1-r^{N_1-k})r^{k}}{\frac{r\alpha}{1+r\alpha}+\frac{(1-r)\alpha}{1+(1-r)\alpha}}}_{\displaystyle{\rm first \ part}}\underbrace{\displaystyle+\frac{(1-r^{N_1})L}{1}+\displaystyle\frac{r^{N_1}L}{1+p}}_{\displaystyle{\rm second \ part}} & \ \ {\rm for} \ r\neq 1 \\
%\\
%\underbrace{\displaystyle\frac{1}{\alpha}+\displaystyle\frac{N-1}{\frac{\alpha}{1+\alpha}}}_{\displaystyle{\rm first \ part}}\underbrace{+\frac{L}{1+p}}_{\displaystyle{\rm second \ part}} & \ \ {\rm for} \ r= 1\\
%\end{array} \right.
\end{split}
\label{eq:TSW2}
\end{eqnarray}
\flushleft
\begin{equation}
\begin{split}
T_{\rm SW}&>\displaystyle\frac{1}{\alpha}+\displaystyle\sum_{k=1}^{N_1} \{P(k)\times(T_{\rm SW}^{\rm s}(N-N_1)+T_{\rm S}^{\rm f})\} + \left(1-\sum_{k=1}^{N_1} P(k)\right)\times(T_{\rm SW}^{\rm s}(N)+T_{\rm W}^{\rm f}) 
\\
&=\displaystyle\frac{1}{\alpha}+\displaystyle\sum_{k=1}^{N_1} P(k) \times \left(\frac{N-1}{Q_{\rm SW}}-\frac{N_1}{Q_{\rm SW}}+\frac{L}{1}\right) + \left(1-\sum_{k=1}^{N_1} P(k)\right)\times\left(\frac{N-1}{Q_{\rm SW}}+\frac{L}{1+p}\right) 
\\
&\geq\displaystyle\frac{1}{\alpha}+\displaystyle\sum_{k=1}^{N_1} P(k) \times \left(\frac{N-1}{Q_{\rm SW}}-\frac{N_0}{Q_{\rm SW}}+\frac{L}{1}\right) + \left(1-\sum_{k=1}^{N_1} P(k)\right)\times\left(\frac{N-1}{Q_{\rm SW}}+\frac{L}{1+p}\right) 
\\
&=\displaystyle\frac{1}{\alpha}+\displaystyle\frac{N-1}{Q_{\rm SW}}+\displaystyle\frac{L}{1+p}\geq\displaystyle\frac{1}{\alpha}+\displaystyle\frac{N-1}{Q_{\rm WW}}+\displaystyle\frac{L}{1+p}=T_{\rm WW}
\end{split}
\label{eq:TSW3}
\end{equation}
\flushleft
\begin{equation}
\begin{split}
T_{\rm SW}-T_{\rm SS}&=\displaystyle\frac{1}{\alpha}+\displaystyle\frac{N-1}{Q_{\rm SW}}-\displaystyle\frac{\sum_{k=1}^{N_1}(1-r^{N_1-k})r^{k}}{Q_{\rm SW}}+\displaystyle\frac{(1-r^{N_1})L}{1}+\displaystyle\frac{r^{N_1}L}{1+p}-\left(\frac{N}{Q_{\rm SS}}+\frac{L}{1}\right)\\
&=\underbrace{(N-1)\left(\frac{1}{Q_{\rm SW}}-\frac{1}{Q_{\rm SS}}\right)-\displaystyle\frac{\sum_{k=1}^{N_1}(1-r^{N_1-k})r^{k}}{Q_{\rm SW}}}_{\displaystyle{\rm first \ part}}\underbrace{-\displaystyle\frac{pr^{N_1}}{1+p}\frac{L}{1}}_{\displaystyle{\rm second \ part}}
\end{split}
\label{eq:TSWTSS}
\end{equation}
\end{figure*}

Similarly to the assumptions as those applied to $T_{\rm SS}$ and $T_{\rm WW}$, if we assume that (iii) particles tend to enter the system every $1/Q_{\rm SW}$ time steps, and that (iv) a standing-preference (walking-preference) particle on average stays in the system for $L/1$ ($L/(1+p)$) time steps, $N_0$ satisfies
\setcounter{equation}{11}
\begin{equation}
(N_0-1)\times\frac{1}{Q_{\rm SW}}\approx \frac{L}{1}-\frac{L}{1+p}.
\end{equation}
Consequently, $N_0$ reduces to 
\begin{equation}
\begin{split}
N_0&\approx \left(\frac{L}{1}-\frac{L}{1+p}\right)\frac{1}{Q_{\rm SW}}+1\\
&=\frac{pL}{1+p}\left\{\frac{r\alpha}{1+r\alpha}+\frac{(1-r)\alpha}{1+(1-r)\alpha}\right\}+1.
\end{split}
\label{eq:N0}
\end{equation}
We note that although $N_0$ is defined to be consecutive, we use its integer part when it is used as the upper limit of summation, which must be an integer.

Regarding $T_{\rm SW}$ as a random variable, i.e., $T_{\rm SW}=t_{\rm SW}(k)$ ($k=1,2,......,N+1$), $t_{\rm SW}(k)$ can be represented as
\begin{eqnarray}
\begin{split}
&t_{\rm SW}(k)\\
&\approx\left\{ \begin{array}{ll}
\displaystyle\frac{1}{\alpha}+T_{\rm SW}^{\rm s}(N-k+1)+T_{\rm S}^{\rm f} & {\rm for} \ 1\leq k<N_1, \\
\\
\displaystyle\frac{1}{\alpha}+T_{\rm SW}^{\rm s}(N)+T_{\rm W}^{\rm f} & {\rm for} \ N_1 \leq k\leq N+1,
\end{array} \right.
\end{split}
\label{eq:TSW1}
\end{eqnarray}
where $N_1={\rm min}(N, N_0)$.

\begin{figure*}[htbp]
\begin{center}
\includegraphics[width=17.5cm,clip]{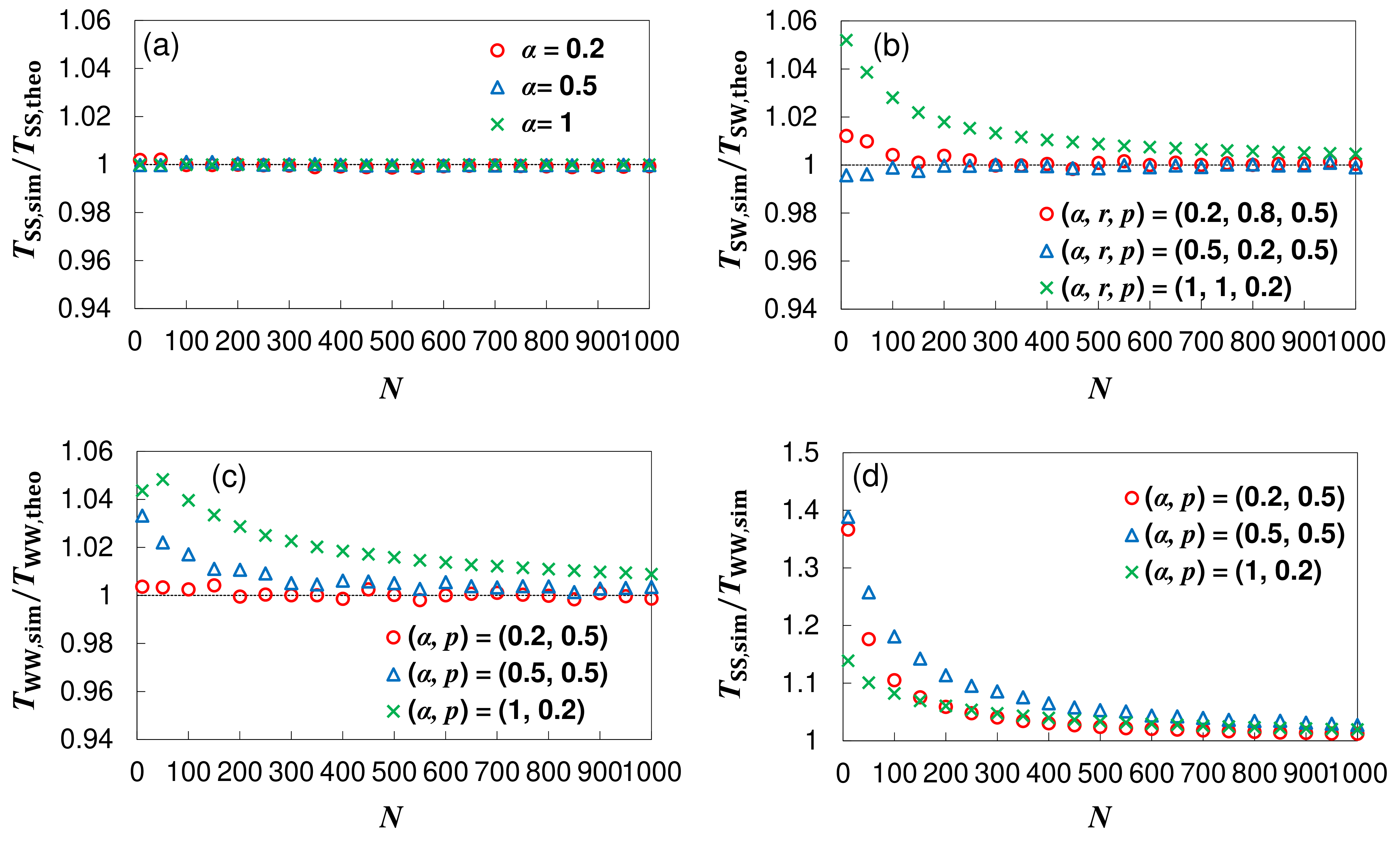}
\caption{(Color Online) Calculated values of the ratios (a) $T_{\rm SS,sim}/T_{\rm SS,theo}$ for $\alpha\in\{0.2  {\rm (red \ circles)},0.5  {\rm (blue \ triangles)} ,1  {\rm (green \ crosses)}\}$, (b) $T_{\rm SW,sim}/T_{\rm SW,theo}$ for various $(\alpha,r,p)\in\{(0.2,0.8,1)  {\rm (red \ circles)}, (0.5,0.5,0.5)  {\rm (blue \ triangles)}, (1,1,0.2)  {\rm (green \ crosses)}\}$, (c) $T_{\rm WW,sim}/T_{\rm WW,sim}$ for various $(\alpha,p)\in\{(0.2,1)  {\rm (red \ circles)}, (0.5,0.5)  {\rm (blue \ triangles)}, (1,0.2)  {\rm (green \ crosses)}\}$, (d) $T_{\rm SS,sim}/T_{\rm WW,theo}$ for various $(\alpha,p)\in\{(0.2,1)  {\rm (red \ circles)}, (0.5,0.5)  {\rm (blue \ triangles)}, (1,0.2)  {\rm (green \ crosses)}\}$. All the plots start from $N=10$.}
\label{fig:validityT}
\end{center}
\end{figure*}

\setcounter{equation}{20}
\begin{figure*}
\flushleft
\begin{equation}
\lim_{N \to \infty} \displaystyle\frac{T_{\rm SW}}{T_{\rm SS}}=\lim_{N \to \infty} \displaystyle\frac{\frac{1}{\alpha}+\frac{N-1}{Q_{\rm SW}}-\frac{\sum_{k=1}^{N_1}(1-r^{N_1-k})r^{k}}{Q_{\rm SW}}+\frac{(1-r^{N_1})L}{1}+\frac{r^{N_1}L}{1+p}}{\frac{N}{Q_{\rm SS}}+\frac{L}{1}}=\lim_{N \to \infty} \displaystyle\frac{\frac{N-1}{Q_{\rm SW}}}{\frac{N}{Q_{\rm SS}}}=\frac{Q_{\rm SS}}{Q_{\rm SW}}>1
\label{eq:TSW/TSS}
\end{equation}
\end{figure*}

\setcounter{equation}{14}

Because each particle prefers standing (walking) with probability $1-r$ ($r$), the probability $P(k)$ that the $k$th-entering particle from the final one is approximately identical to the final-entering standing-preference particle can be calculated as 

\begin{eqnarray}
P(k)=\left\{ \begin{array}{ll}
(1-r)r^{k-1} & {\rm for} \ 1\leq k\leq N, \\
r^{k-1} & {\rm for} \ k=N+1,
\end{array} \right.
\label{eq:prob}
\end{eqnarray}
satisfying $\sum_{k=1}^{N+1} P(k)=1$.

\setcounter{equation}{18}

Using Eqs. (\ref{eq:TSW1}) and (\ref{eq:prob}), the expected value of $T_{\rm SW}$ can be calculated as Eq. (\ref{eq:TSW2}). The detailed derivation of $T_{\rm SW}$ is given in Appendix \ref{sec:TSW}.

Next, we compare $T_{\rm SS}$, $T_{\rm SW}$, and $T_{\rm WW}$, using Eqs. (\ref{eq:TSS}), (\ref{eq:TWW}), and (\ref{eq:TSW2}). First, $T_{\rm SS}$ clearly exceeds $T_{\rm WW}$. In addition, because of Eq. (\ref{eq:TSW3}), $T_{\rm SW}$ also exceeds $T_{\rm WW}$. 

The relation between $T_{\rm SS}$ and $T_{\rm SW}$ is somewhat complicated. Using Eqs. (\ref{eq:TSS}) and (\ref{eq:TSW2}), $T_{\rm SW}-T_{\rm SS}$ can be calculated as in Eq. (\ref{eq:TSWTSS}).

First, the second part of Eq. (\ref{eq:TSWTSS}) is always equal to or less than 0 (the equal sign holds only when $r=0$), and can be regarded as a function of $L$.
This is due to the shorter dwell time of walking-preference particles compared to standing-preference particles (see also Fig. \ref{fig:dwelltime}). We hereafter refer to this effect as the `positive effect of walking,' and it disappears gradually as $r$ decreases or $N_1$ increases, since $r^{N_1}\approx 0$.

On the other hand, the sign of the first part of Eq. (\ref{eq:TSWTSS}) depends on the parameters.
For $N>1$, the first term in the first part is always positive, due to $Q_{\rm SS}>Q_{\rm SW}$ (see Eq. (\ref{eq:QSWQSS})), and can be regarded as a function of $N$.

However, the existence of the second term in the first part of Eq. (\ref{eq:TSWTSS}) invites the result that the first part of Eq. (\ref{eq:TSWTSS}) becomes negative. The numerator of this term satisfies the following relation:
\begin{equation}
\sum_{k=1}^{N_1}(1-r^{N_1-k})r^{k}<N_1(1-r^{N_1})<N_1.
\end{equation}
From the definition of $N_1$, this numerator increases monotonically and converges to a constant value; specifically, $\sum_{k=1}^{N_0}(1-r^{N_0-k})r^{k}$ when $N\geq N_0$ , as $N$ increases. From Eq. (\ref{eq:N0}), this constant value can be regarded as a monotonically increasing function of $L$.

Therefore, for sufficiently large $N$, the first part of Eq. (\ref{eq:TSWTSS}) exceeds 0. We hereafter refer this effect as the `negative effect of preference,' because a difference arises due to $Q_{\rm SS}>Q_{\rm SW}$, which is introduced by the walking/standing preference (see Eq. (\ref{eq:QSWQSS})). 

Consequently, $T_{\rm SW}-T_{\rm SS}$ for $N>N_0$ can be rewritten as
\begin{equation}
T_{\rm SW}-T_{\rm SS}=f(N)-g(L),
\label{eq:differ}
\end{equation} 
where $f(N)$ and $g(L)$ are, respectively, monotonically increasing functions of $N$ and $L$. This fact implies that for sufficiently large $N$ the positive effect of walking becomes negligible and $T_{\rm SW}-T_{\rm SS}>0$ ($T_{\rm SW}>T_{\rm SS}$), while this relation can be reversed for small $N$.
From Eqs. (\ref{eq:TSS}) and (\ref{eq:TSW2}), $T_{\rm SW}/T_{\rm SS}$ converges to a certain value as $N$ increases, as in Eq. (\ref{eq:TSW/TSS}).

The discussion in this subsection indicates that (i) Strategy WW is always advantageous over Strategy SS and SW, and that (ii) Strategy SS is advantageous over Strategy SW if $N$ is sufficiently large; otherwise Strategy SW performs better than Strategy SS.

\subsubsection{Validation of approximations of $T$}
\label{sec:validityT}
This subsection discusses the validity of the theoretical approximations of $T_{\rm SS}$, $T_{\rm SW}$, and $T_{\rm WW}$ by comparing them with the results of the numerical simulations. 

Figure \ref{fig:validityT} plots the calculated values of the ratios (a) $T_{\rm SS,sim}/T_{\rm SS,theo}$ for various $\alpha\in\{0.2,0.5 ,1\}$, (b) $T_{\rm SW,sim}/T_{\rm SW,theo}$ for various $(\alpha,r,p)\in\{(0.2,0.8,1), (0.5,0.5,0.5), (1,1,0.2)\}$, (c) $T_{\rm WW,sim}/T_{\rm WW,sim}$ for various $(\alpha,p)\in\{(0.2,1), (0.5,0.5), (1,0.2)\}$, and (d) $T_{\rm SS,sim}/T_{\rm WW,theo}$ for various $(\alpha,p)\in\{(0.2,1), (0.5,0.5), (1,0.2)\}$. We average the simulation values of $T$ over 1000 trials (and do similarly below for the simulations of $T$ unless otherwise specified). 

Figure \ref{fig:validityT} (a) shows that the simulation values of $T_{\rm SS}$ agree with the theoretical ones very well, even for small $N$. Conversely, Figs. \ref{fig:validityT} (b) and (c) show that the simulation values of $T_{\rm SW}$ and $T_{\rm WW}$ agree with the theoretical ones very well for large $N$; however, the simulation diverges from the analysis for small $N$ and especially if $p$ is small. This behavior can be explained as follows.

From Eq. (\ref{eq:Tf}), $T_{\rm S}^{\rm f}$ is deterministic, whereas $T_{\rm W}^{\rm f}$ is expected. In addition, although the final-leaving particle is assumed to be able to hop forward freely in the theoretical calculations, it may be blocked by the particle ahead of it in a walking lane, so $T_{\rm W}^{\rm f}>L/(1+p)$. These features mean that $T_{\rm SW}$ and $T_{\rm WW}$ can diverge in the theoretical approximations, especially for small $N$ and small $p$, as these conditions enhance the influence of $T_{\rm W}^{\rm f}$ on $T_{\rm SW}$ and $T_{\rm WW}$.

In Fig. \ref{fig:validityT} (d), $T_{\rm SS}$ is confirmed to approach $T_{\rm WW}$ as $N$ increases (strictly speaking, $T_{\rm WW}>T_{\rm SS}$). From here on, we do not consider $T_{\rm WW}$, since $T_{\rm WW}<T_{\rm SS}$ and $T_{\rm WW}<T_{\rm SW}$, because of the difference in $T^{\rm f}$.

\subsection{Effect of $N$}
\label{sec:Ndepend}
In this subsection, we investigate the effect of $N$ on $T$ for various $r$. We set $\alpha=0.5$ and $p=0.5$ (in a walking lane). 

\begin{figure}[htbp]
\begin{center}
\includegraphics[width=8.5cm,clip]{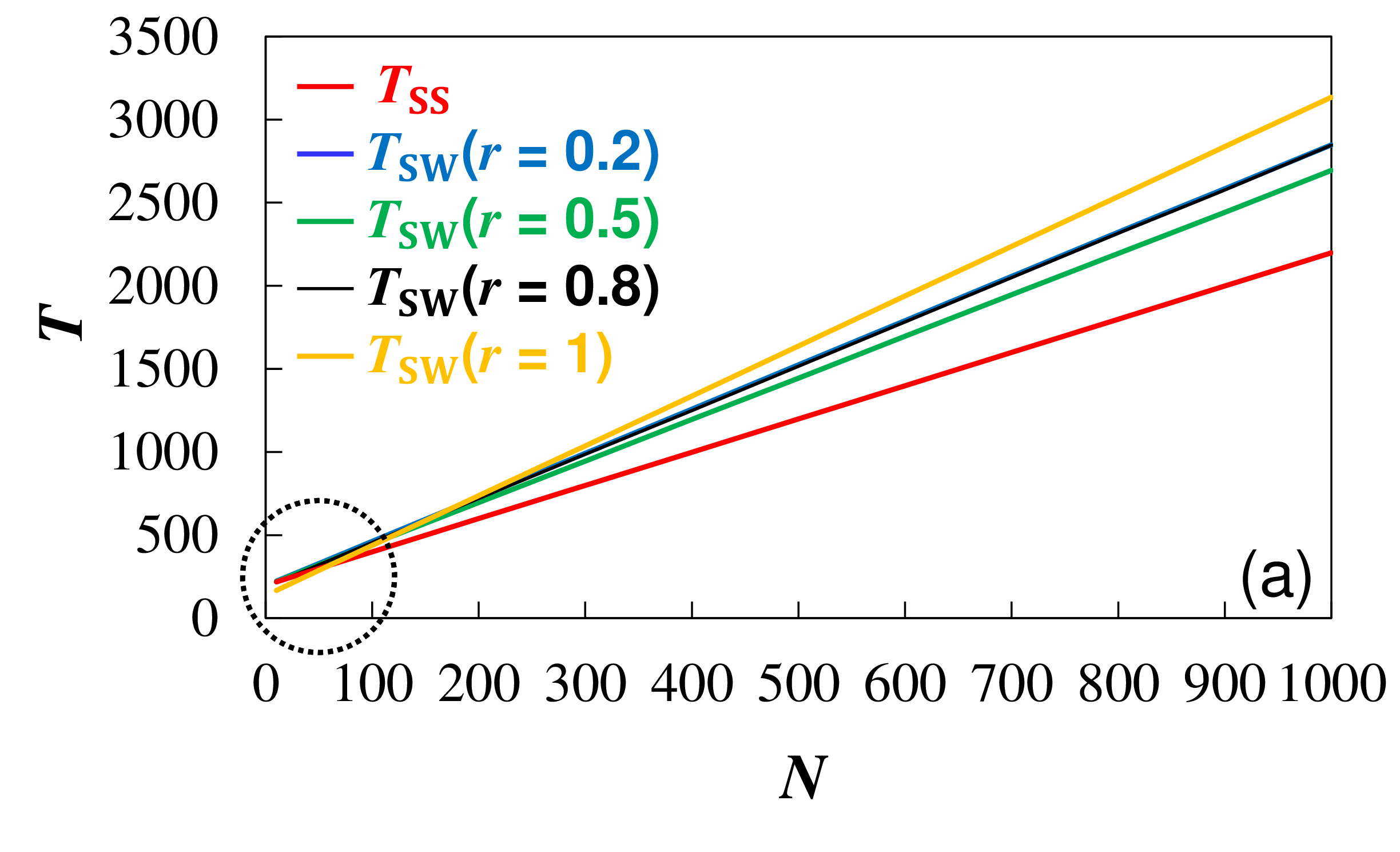}\\
\includegraphics[width=8.5cm,clip]{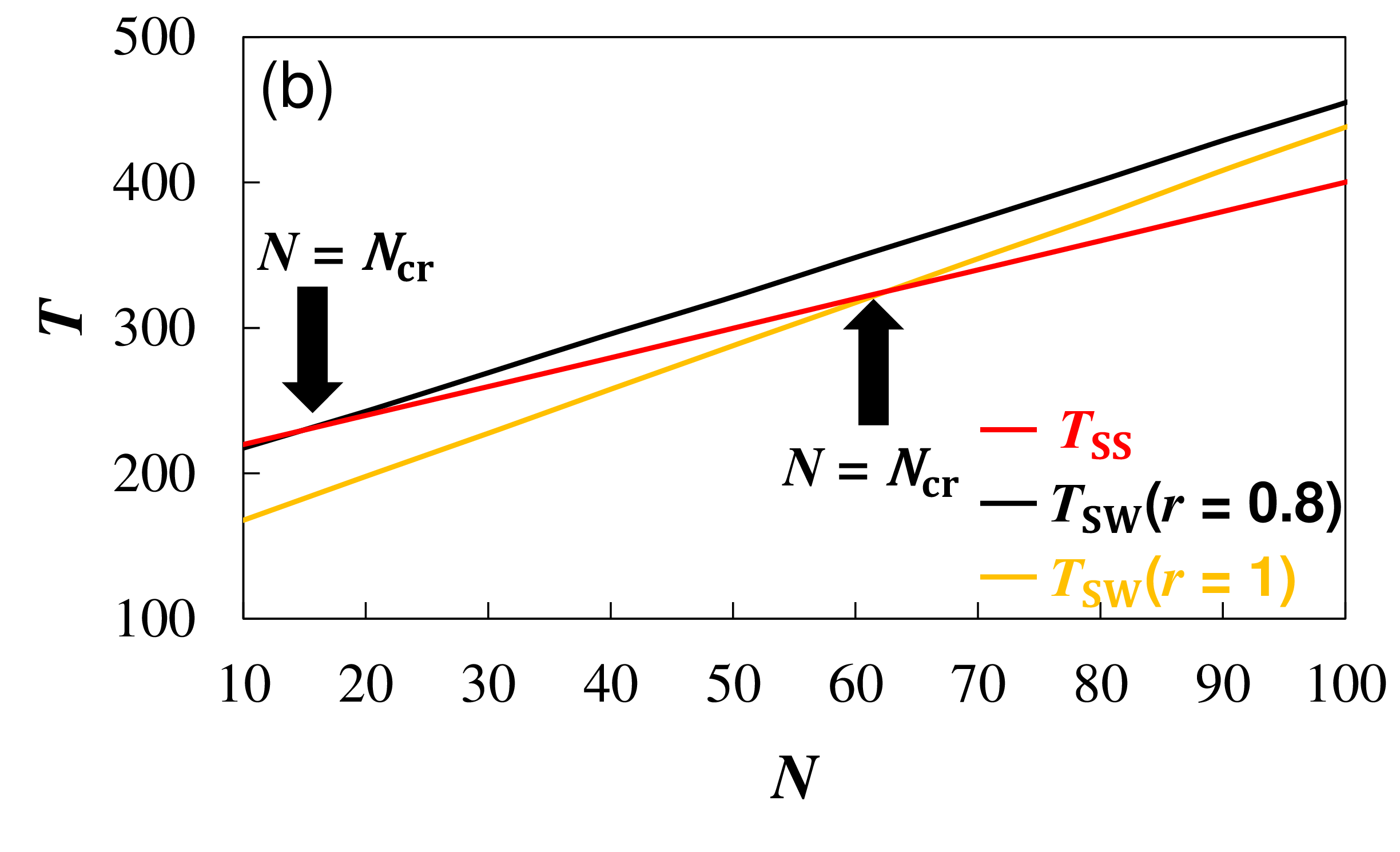}
\caption{(Color Online) (a) Simulation values of $T_{\rm SS}$ (red) and $T_{\rm SW}$ for various $r\in\{0.2 {\rm (blue)}, 0.5  {\rm (green)}, 0.8  {\rm (black)},1  {\rm (yellow)}\}$ as functions of $N$, fixing $\alpha=0.5$ and $p=0.5$. All the plots start from $N=10$. (b) Zoomed Fig. \ref{fig:Ndepend} (a), in the area enclosed with a black dotted circle in Fig. \ref{fig:Ndepend} (a). We note that the curves for $r=0.2$ and $r=0.5$ are abbreviated. We visually mark the vicinities of $N=N_{\rm cr}$, which will be discussed in Subsec. \ref{sec:Ncrrcr}, with black arrows.}
\label{fig:Ndepend}
\end{center}
\end{figure}

Figure \ref{fig:Ndepend} (a) compares $T_{\rm SS}$ and $T_{\rm SW}$ for various $r\in\{0.2,0.5,0.8,1\}$ as functions of $N$ and Fig. \ref{fig:Ndepend} (b) is a zoomed-in inset of Fig. \ref{fig:Ndepend} (a).

In Fig. \ref{fig:Ndepend} (a), we can observe that $T_{\rm SW}$ generally exceeds $T_{\rm SS}$ at least for $N>100$. This trend means that the negative effect of preference is always greater than the positive effect of walking for large $N$. Moreover, $T_{\rm SW}-T_{\rm SS}$ increases as $N$ becomes greater. This phenomenon is expected from Eq. (\ref{eq:differ}). We note that the curves for $r=0.2$ and $r=0.8$ mostly overlap, even though they are, strictly speaking, different (see also Eq. (\ref{eq:TSW2}))---because the steady-state flows are the same.

On the other hand, in Fig. \ref{fig:Ndepend} (b), some cases of $T_{\rm SS}>T_{\rm SW}$ appear for small $N$, especially for relatively large $r$. This behavior is because in those limited cases, $T_{\rm SW}$ is influenced very much by the positive effect of walking, i.e., $r^{N_1}$ is relatively large.

\setcounter{equation}{24}
\begin{figure*}
\flushleft
\begin{equation}
\begin{split}
\lim_{\alpha \to 0} \displaystyle\frac{T_{\rm SW}}{T_{\rm SS}}&=\lim_{\alpha \to 0} \displaystyle\frac{\frac{1}{\alpha}+\frac{N-1}{Q_{\rm SW}}-\frac{\sum_{k=1}^{N_1}(1-r^{N_1-k})r^{k}}{Q_{\rm SW}}+\frac{(1-r^{N_1})L}{1}+\frac{r^{N_1}L}{1+p}}{\frac{N}{Q_{\rm SS}}+\frac{L}{1}}\\
&=\lim_{\alpha \to 0} \displaystyle\frac{N-\sum_{k=1}^{N_1}(1-r^{N_1-k})r^{k}+\frac{\alpha(1-r^{N_1})L}{1}+\frac{\alpha r^{N_1}L}{1+p}}{N+\frac{\alpha L}{1}}=\lim_{\alpha \to 0}\left\{1-\frac{\sum_{k=1}^{N_1}(1-r^{N_1-k})r^{k}}{N}\right\}=1
\end{split}
\label{eq:TSWTSSalpha}
\end{equation}
\end{figure*}

\subsection{Effect of $\alpha$}
\label{sec:alphadepend}
In this subsection, we investigate the effect of $\alpha$ on $T$ for various $N$. We set $r=0.5$, and $p=0.5$ (in a walking lane).

\begin{figure}[htbp]
\begin{center}
\includegraphics[width=8.5cm,clip]{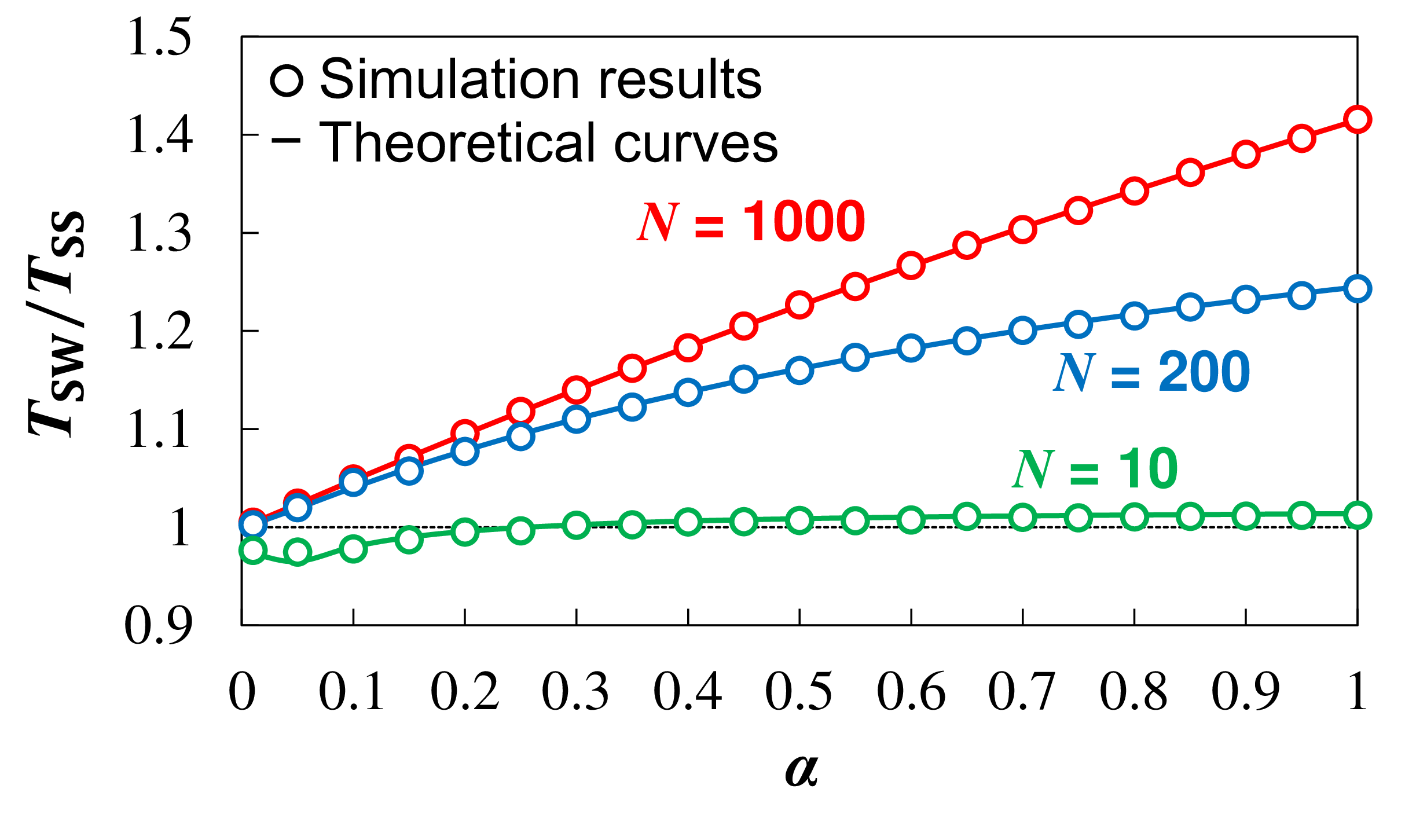}
\caption{(Color Online) Simulation (circles) and theoretical (curves) values of the ratio $T_{\rm SW}/T_{\rm SS}$ as functions of $\alpha$ for various $N\in\{10 ({\rm green}), 200 ({\rm blue}),1000 ({\rm red})\}$. The other parameters are fixed as $r=0.5$ and $p=0.5$, respectively. All the plots start from $\alpha=0.01$.}
\label{fig:alphadepend}
\end{center}
\end{figure}

Figure \ref{fig:alphadepend} compares the simulation (circles) and theoretical (curves) values for the ratio $T_{\rm SW}/T_{\rm SS}$ as functions of $\alpha$ for various $N\in\{10, 200,1000\}$. The simulation results show very good agreement with the theoretical curves. The inequality $T_{\rm SW}/T_{\rm SS}>1$ ($T_{\rm SW}/T_{\rm SS}<1$) means that Strategy SS (Strategy SW) is advantageous. 

In Fig. \ref{fig:alphadepend}, again we see that $T_{\rm SW}/T_{\rm SS}>1$ ($T_{\rm SS}<T_{\rm SW}$) regardless of $\alpha$ for large $N$ ($N=200$ and 1000 in this figure). 

On the other hand, $T_{\rm SW}/T_{\rm SS}$ decreases as $\alpha$ decreases, and $T_{\rm SW}/T_{\rm SS}$ can become less than 1 ($T_{\rm SS}>T_{\rm SW}$) for small $\alpha$ and small $N$ ($N=10$ and $\alpha<0.2$ in this figure). This trend is explained as follows.

\setcounter{equation}{21}
From Eqs. (\ref{eq:QSS}), (\ref{eq:QSW}), and (\ref{eq:TSW/TSS}) we obtain 
\begin{equation}
\lim_{N\to \infty}\frac{T_{\rm SW}}{T_{\rm SS}}=\displaystyle\frac{Q_{\rm SS}}{Q_{\rm SW}}=\frac{1}{\frac{r}{1+r\alpha}+\frac{1-r}{1+(1-r)\alpha}},
\label{eq:limit}
\end{equation}
which increases monotonically with respect to $\alpha$. Therefore, for sufficiently large $N$, $T_{\rm SW}/T_{\rm SS}$ increases (decreases) as $\alpha$ increases (decreases). 

Then, taking note of the following two relations: 
\begin{equation}
\lim_{\alpha \to 0} \displaystyle\frac{Q_{\rm SW}}{Q_{\rm SS}}=\lim_{\alpha \to 0}\left\{\frac{r}{1+r\alpha}+\frac{1-r}{1+(1-r)\alpha}\right\}=1
\label{eq:QSWQSSalpha}
\end{equation}
and
\begin{equation}
\begin{split}
&\lim_{\alpha \to 0} N_1=\lim_{\alpha \to 0} N_0\\
&=\lim_{\alpha \to 0} \left[\frac{pL}{1+p}\left\{\frac{r\alpha}{1+r\alpha}+\frac{(1-r)\alpha}{1+(1-r)\alpha}\right\}+1\right]=1,
\end{split}
\label{eq:N1alpha}
\end{equation}
we have Eq. (\ref{eq:TSWTSSalpha}). In Fig. \ref{fig:alphadepend}, all the plots (curves) are observed to approach 1 as $\alpha$ approaches 0. Eq. (\ref{eq:QSWQSSalpha}) implies that for small $\alpha$, the negative effect of preference disappears. 

Here, we define a new value $\alpha=\alpha_1$, where $\alpha_1$ ($0<\alpha_1\leq1$) satisfies
\setcounter{equation}{25}
\begin{equation}
\begin{split}
&N_0(\alpha=\alpha_1)=2\\
&\Leftrightarrow\frac{pL}{1+p}\left\{\frac{r\alpha_1}{1+r\alpha_1}+\frac{(1-r)\alpha_1}{1+(1-r)\alpha_1}\right\}=1.
\end{split}
\end{equation}

For sufficiently large $N$, $T_{\rm SW}-T_{\rm SS}$ always exceeds 0 when $\alpha_1<\alpha\leq1$, which is discussed in detail in Appendix \ref{sec:monotonicity}.

On the other hand, when $0<\alpha\leq\alpha_1$, from Eq. (\ref{eq:TSWTSS}),
\begin{equation}
\begin{split}
&T_{\rm SW}-T_{\rm SS}\\
&=(N-1)\left(\frac{1}{Q_{\rm SW}}-\frac{1}{Q_{\rm SS}}\right)-\displaystyle\frac{rp}{1+p}\frac{L}{1}\\
%&=\displaystyle\frac{N-1}{Q_{\rm SW}}\left\{1-\displaystyle\frac{Q_{\rm SW}}{Q_{\rm SS}}-\displaystyle\frac{rpL}{(1+p)(N-1)}\right\}\\
&=\displaystyle\frac{N-1}{Q_{\rm SW}}\times g(\alpha)>0,
\end{split}
\label{eq:TSWTSSg}
\end{equation}
where $g(\alpha)$ is defined as
\begin{equation}
g(\alpha)=1-\displaystyle\frac{Q_{\rm SW}}{Q_{\rm SS}}-\displaystyle\frac{rpL}{(1+p)(N-1)}.
\end{equation}
The behavior of $g(\alpha)$ is discussed in detail in Appendix \ref{sec:galpha}. We note that $N_1=N_0=1$ for the upper limit of summation when $\alpha\leq\alpha_1$ because we need to use the integer part for that case. 
Eq. (\ref{eq:TSWTSSg}) indicates that for sufficiently large $N$; especially, 
\begin{equation}
N>\frac{pL}{1+p}\max\left(\frac{r}{1-r},1\right)+1,
\end{equation}
$T_{\rm SW}/T_{\rm SS}$ always exceeds 1, and converges to 1 with $\alpha \to 0$, because the negative effect of preference is always greater than the positive effect of walking when $0<\alpha\leq1$.

On the other hand, for small $N$, as $\alpha$ decreases, $T_{\rm SW}/T_{\rm SS}$ can become below 1, because the positive effect of walking is greater than the negative effect of walking, and finally $T_{\rm SW}/T_{\rm SS}$ converges to 1 from Eq. (\ref{eq:TSWTSSalpha}). 
 
\subsection{Effect of $r$}
\label{sec:rdepend}
In this subsection, we investigate the effect of $r$ on $T_{\rm SW}$ for various $N$. We set $\alpha=0.5$ and $p=0.5$ (in a walking lane). We note that only the standing (walking) lane is used and the other lane is always vacant if $r=0$ ($r=1$).

\begin{figure}[htbp]
\begin{center}
\includegraphics[width=8.5cm,clip]{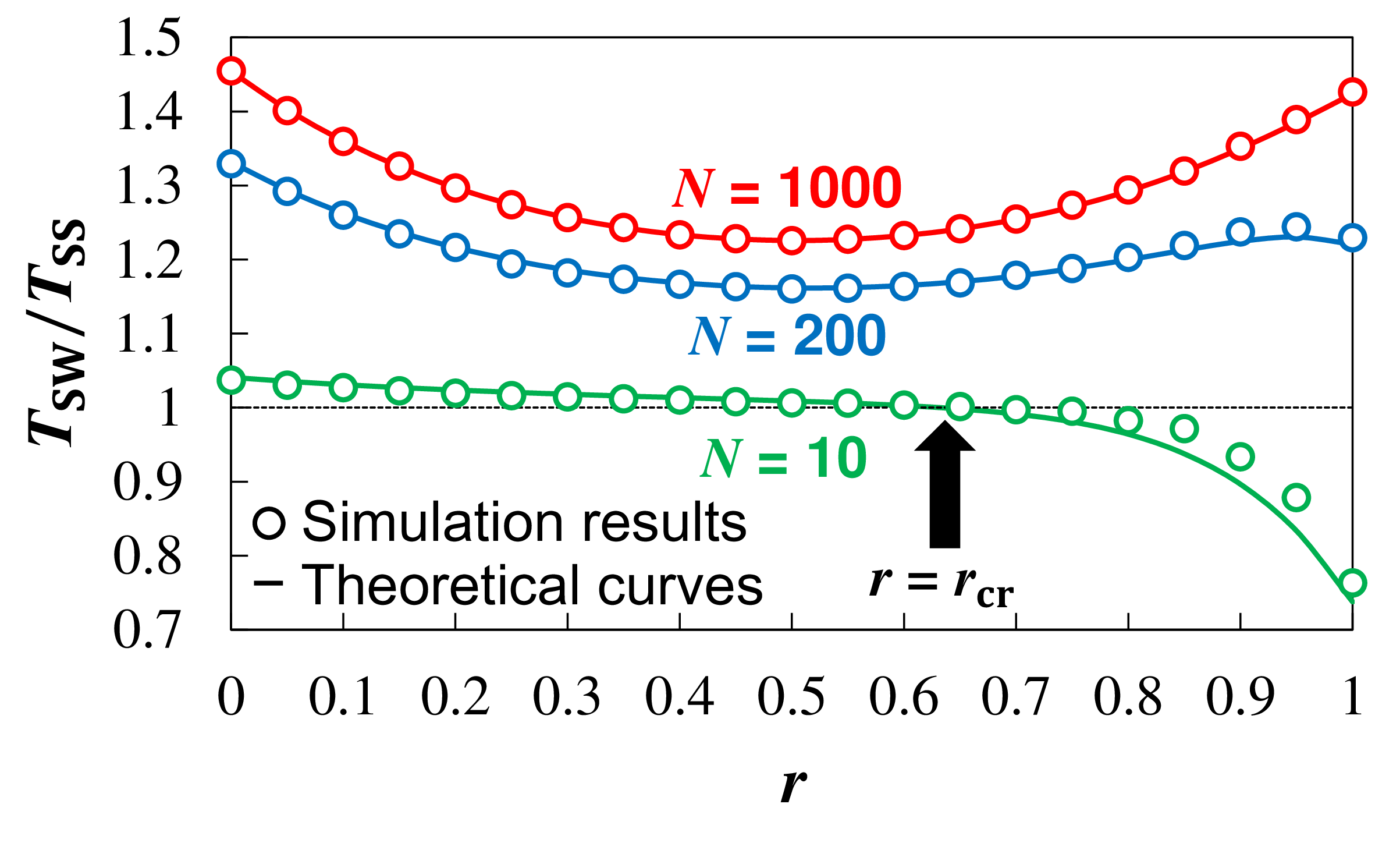}
\caption{(Color Online) Simulation (circles) and theoretical (curves) values of the ratio $T_{\rm SW}/T_{\rm SS}$ as functions of $r$ for various $N\in\{10 ({\rm green}), 200 ({\rm blue}),1000 ({\rm red})\}$. The other parameters are fixed as $\alpha=0.5$ and $p=0.5$, respectively. We visually mark the vicinities of $r=r_{\rm cr}$, which will be discussed in Subsec. \ref{sec:Ncrrcr}, with black arrows.}
\label{fig:r}
\end{center}
\end{figure}

Figure \ref{fig:r} compares the simulation (circles) and theoretical (curves) values of the ratio $T_{\rm SW}/T_{\rm SS}$ as functions of $r$ for various $N\in\{10 , 200 ,1000\}$. We emphasize that $T_{\rm SS}$ is constant for all values of $r$. The simulation results again show very good agreement with the theoretical curves even though they diverge in very limited cases of small $N$ and large $r$ ($N=10$ and $r>0.7$ in this figure). 

This divergence can be explained as follows. First, the positive effect of walking is dominant for small $N$ and large $r$, since the influence of the negative effect of preference is small and $r^{N_1}$ approaches 1. In addition, although the final leaving particle is assumed to be able to hop forward freely in the theoretical calculations, it can be blocked by the particle ahead of it in a walking lane, leading to a slight underestimation of $T_{\rm SW}$ (see Subsec. \ref{sec:validityT}). Consequently, small $N$ and large $r$ enhance the influence of $L/(1+p)$ and increase the likelihood of underestimating it when calculating the theoretical values of $T_{\rm SW}$, so the theoretical curve remains below the simulation values. 

In Fig. \ref{fig:r}, we see that $T_{\rm SW}/T_{\rm SS}>1$ ($T_{\rm SS}<T_{\rm SW}$) for large $N$ ($N=200$ and 1000 in this figure) regardless of $r$, whereas $T_{\rm SW}/T_{\rm SS}<1$ ($T_{\rm SS}>T_{\rm SW}$) for small $N$ and large $r$ ($N=10$ and $r>0.7$ in this figure).

In addition, we also note that $T_{\rm SW}/T_{\rm SS}$ takes its minimum value near $r=0.5$ for sufficiently large $N$ ($N=200$ and $N=1000$ in this figure), because the negative effect of preference is least active with $r$ being slightly more than 0.5, due to the positive effect of walking. 

On the other hand, $T_{\rm SW}/T_{\rm SS}$ can take a minimum value with $r=1$ for sufficiently small $N$ ($N=10$ in this figure), due to the enhanced positive effect of walking for small $N$ and large $r$. Due to the positive effect of walking, $T_{\rm SW}/T_{\rm SS}(r=0)>T_{\rm SW}/T_{\rm SS}(r=1)$ even for large $N$.

\subsection{Effect of $p$}
\label{sec:pdepend}
In this subsection, we investigate the effect of $p$ (in a walking lane) on $T_{\rm SW}$ for various $N$. We set $\alpha=0.5$, and $r=0.5$. We emphasize that $T_{\rm SS}$ is constant regardless of $p$, as it is when changing $r$.

\begin{figure}[htbp]
\begin{center}
\includegraphics[width=8.5cm,clip]{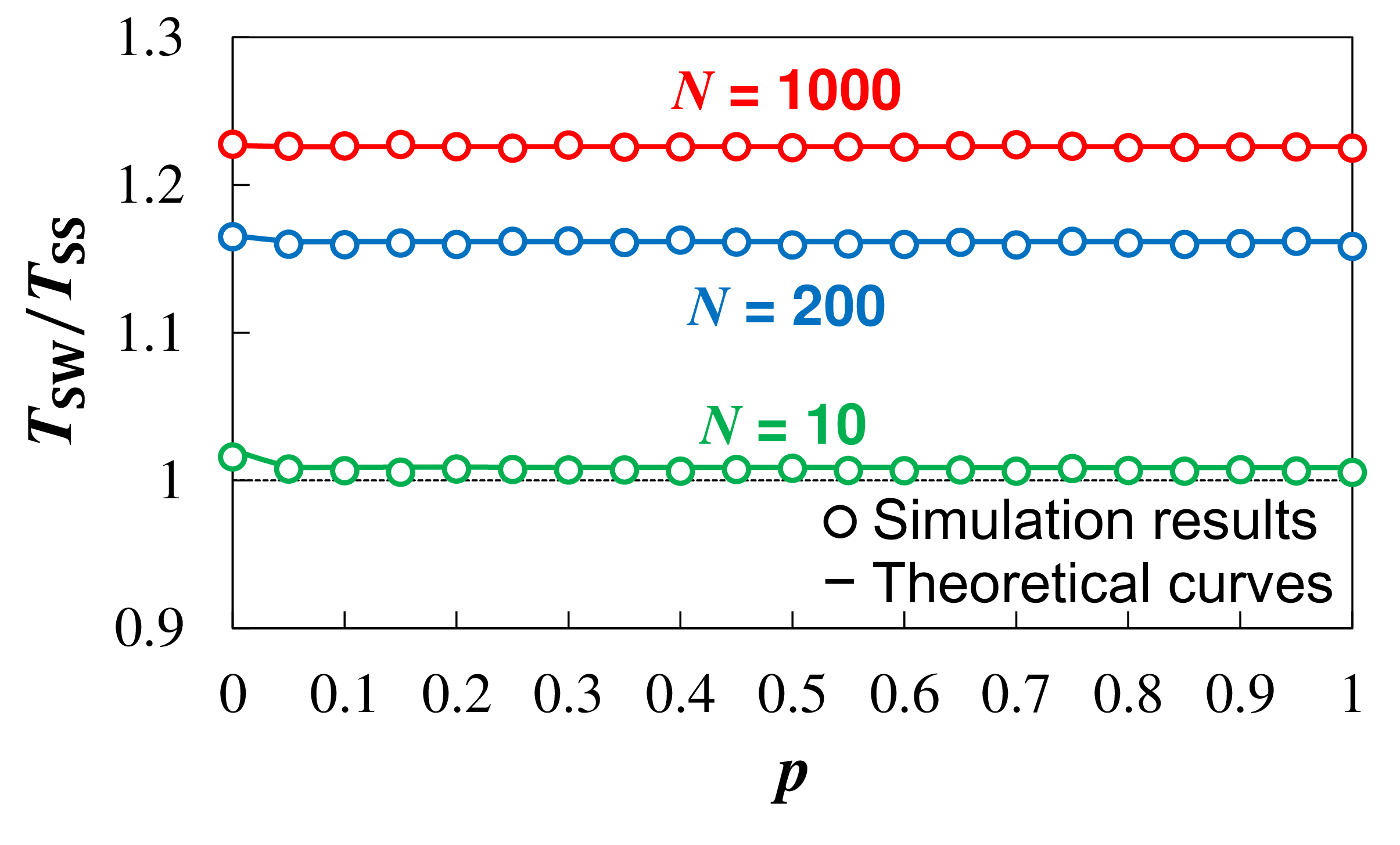}
\caption{(Color Online) Simulation (circles) and theoretical (curves) values of $T_{\rm SW}/T_{\rm SS}$ as a function of $p$ for various $N\in\{50 ({\rm green}), 200 ({\rm blue}),1000 ({\rm red})\}$. The other parameters are fixed as $\alpha=0.5$ and $r=0.5$, respectively.}
\label{fig:p}
\end{center}
\end{figure}

Figure \ref{fig:p} plots the simulation (circles) and theoretical (curves) values of $T_{\rm SW}/T_{\rm SS}$ as functions of $p$ for various $N\in\{10, 200,1000 \}$. The simulation values again show good agreement with the theoretical curves.

In Fig. \ref{fig:p}, we see that $T_{\rm SW}/T_{\rm SS}>1$ ($T_{\rm SS}<T_{\rm SW}$) and $T$ is hardly affected by $p$, because the steady-state flow does not depend on $p$ and the positive effect of walking is very slight when $r=0.5$ and $N\geq10$.  

We note that $T_{\rm SW}/T_{\rm SS}\neq 1$ ($T_{\rm SS}\neq T_{\rm SW}$) with $p=0$, at which both lanes act as standing lanes even when Strategy SW is modeled. In this case, particles can enter either of the two lanes indiscriminately with Strategy SS, while particles still prefer either of the two lanes when Strategy SW is modeled.

\subsection{Reversal point $N_{\rm cr}$ and $r_{\rm cr}$}
\label{sec:Ncrrcr}

\begin{figure*}[htbp]
\begin{center}
\includegraphics[width=17.5cm,clip]{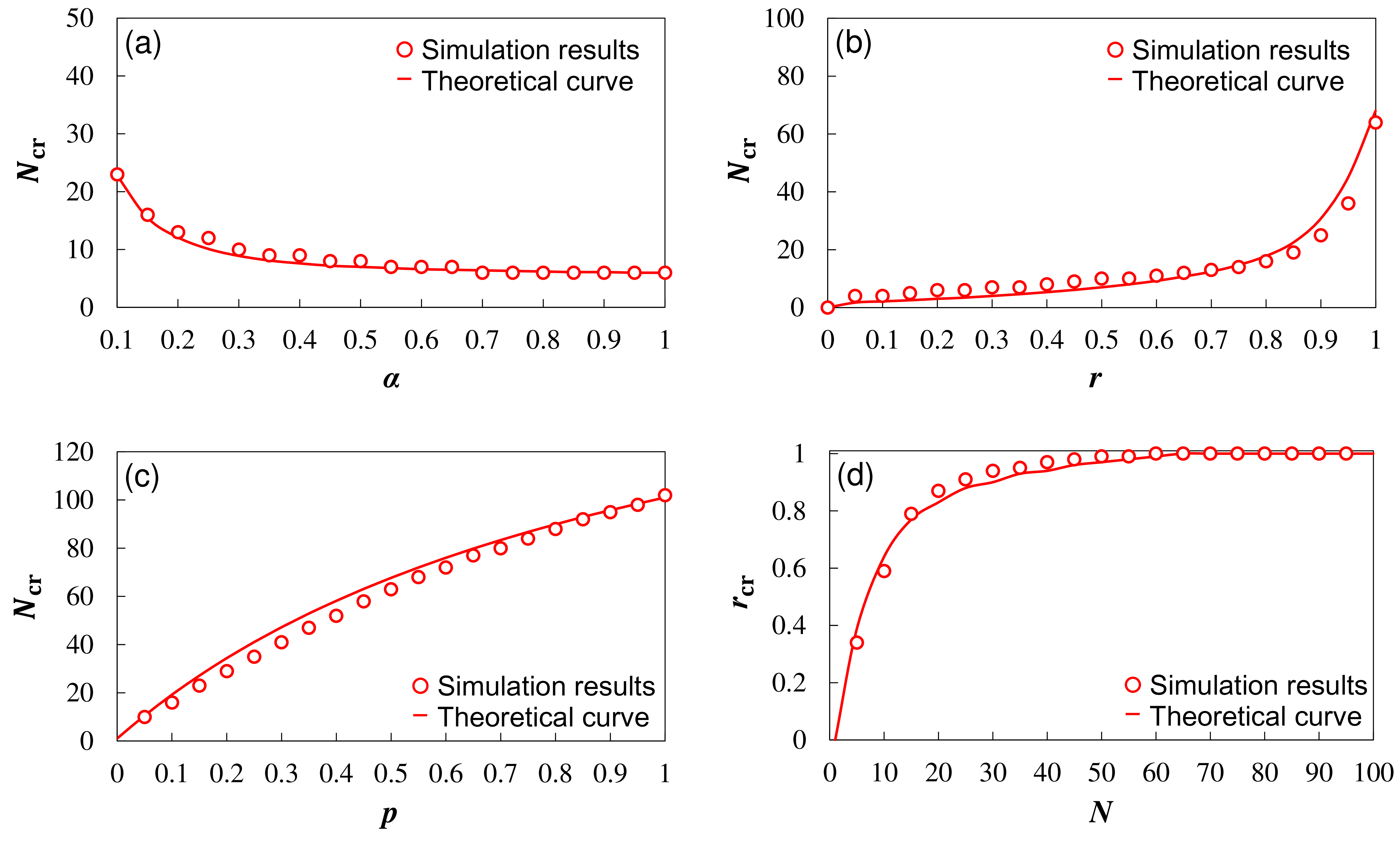}
\caption{(Color Online) Simulation (circles) and theoretical (curve) values of (a) $N_{\rm cr}$ as a function of $\alpha$ with $(r,p)=(0.5,0.5)$, (b) $N_{\rm cr}$ as a function of $r$ with $(\alpha,p)=(0.5,0.5)$, (c) $N_{\rm cr}$ as a function of $p$ with $(\alpha,r)=(0.5,1)$, and (d) $r_{\rm cr}$ as a function of $N$ with $(\alpha,p)=(0.5,0.5)$. We note that the simulation values of $N_{\rm cr}$ satisfy $T_{\rm SS}(N=N_{\rm cr})>T_{\rm SW}(N=N_{\rm cr})$ and $T_{\rm SS}(N=N_{\rm cr}+1)<T_{\rm SW}(N=N_{\rm cr}+1)$, and that those of $r_{\rm cr}$ satisfy $T_{\rm SS}(r=r_{\rm cr})<T_{\rm SW}(r=r_{\rm cr})$ and $T_{\rm SS}(r=r_{\rm cr}+0.01)>T_{\rm SW}(r=r_{\rm cr}+0.01)$, if it exists.}
\label{fig:Ncrrcr}
\end{center}
\end{figure*}

In this subsection, we theoretically determine $N_{\rm cr}(\alpha,r,p)$ and $r_{\rm cr}(N)$, at which $T_{\rm SS}=T_{\rm SW}$ (see also Figs. \ref{fig:Ndepend}---\ref{fig:r}). These theoretical values are then compared with the simulation values. 

From Eqs. (\ref{eq:TSS}) and (\ref{eq:TSW2}), $N_{\rm cr}$ for $0<r<1$ satisfies the following relation:

\begin{equation}
\centering
\begin{split}
&T_{\rm SS}(N=N_{\rm cr})=T_{\rm SW}(N=N_{\rm cr})\\
&\Leftrightarrow\frac{N_{\rm cr}}{\alpha}+\frac{L}{1}\\
&\quad=\displaystyle\frac{1}{\alpha}+\displaystyle\frac{N_{\rm cr}-1}{\frac{r\alpha}{1+r\alpha}+\frac{(1-r)\alpha}{1+(1-r)\alpha}}+\displaystyle\frac{(1-r^{N_1})L}{1}+\displaystyle\frac{r^{N_1}L}{1+p}\\
&\quad\quad-\displaystyle\frac{\frac{r}{1-r}\{1-N_1 r^{N_1-1}+(N_1-1)r^{N_1}\}}{\frac{r\alpha}{1+r\alpha}+\frac{(1-r)\alpha}{1+(1-r)\alpha}},
\end{split}
\label{eq:Ncr}
\end{equation}
where $N_1={\rm min}(N,N_0)={\rm min}(N_{\rm cr},N_0)$.

%\begin{equation}
%T_{\rm SS}(N=N_{\rm cr})=T_{\rm SW}(N=N_{\rm cr})
%\label{eq:Ncr}
%\end{equation}
%We note that for obtaining the theoretical values we regard $N_{\rm cr}$ as a consecutive value for convenience.

$N_{\rm cr}$ is generally difficult to obtain for $0<r<1$; however, for the case $r=1$, the general form of $N_{\rm cr}$ can be obtained easily as
\begin{equation}
\centering
\begin{split}
&T_{\rm SS}(N=N_{\rm cr})=T_{\rm SW}(N=N_{\rm cr})\\
&\Leftrightarrow \frac{N_{\rm cr}}{\alpha}+\frac{L}{1}=\displaystyle\frac{1}{\alpha}+\displaystyle\frac{N_{\rm cr}-1}{\frac{\alpha}{1+\alpha}}+\displaystyle\frac{L}{1+p}\\
&\Leftrightarrow N_{\rm cr}=\displaystyle\frac{pL}{1+p}+1.
\end{split}
\end{equation}

Similarly, $r_{\rm cr}$ ($0<r_{\rm cr}<1$) satisfies the following relation:
\begin{equation}
\centering
\begin{split}
&T_{\rm SS}(r=r_{\rm cr})=T_{\rm SW}(r=r_{\rm cr})\\
&\Leftrightarrow\frac{N}{\alpha}+\frac{L}{1}\\
&\quad=\displaystyle\frac{1}{\alpha}+\displaystyle\frac{N-1}{\frac{r_{\rm cr}\alpha}{1+r_{\rm cr}\alpha}+\frac{(1-r_{\rm cr})\alpha}{1+(1-r_{\rm cr})\alpha}}+\displaystyle\frac{(1-{r_{\rm cr}}^{N_1})L}{1}+\displaystyle\frac{r_{\rm cr}^{N_1}L}{1+p}\\
&\quad\quad-\displaystyle\frac{\frac{r_{\rm cr}}{1-r_{\rm cr}}\{1-N_1 r_{\rm cr}^{N_1-1}+(N_1-1)r_{\rm cr}^{N_1}\}}{\frac{r_{\rm cr}\alpha}{1+r_{\rm cr}\alpha}+\frac{(1-r_{\rm cr})\alpha}{1+(1-r_{\rm cr})\alpha}}.
\end{split}
\label{eq:rcr}
\end{equation}
We note that $r_{\rm cr}=1$ if no value in $0<r_{\rm cr}<1$ that satisfies Eq. (\ref{eq:rcr}) exists, indicating that Strategy SS is always advantageous regardless of $r$ for a fixed $N$.

Figure \ref{fig:Ncrrcr} compares the simulation (circles) and theoretical (curves) values of (a) $N_{\rm cr}$ as a function of $\alpha$, (b) $N_{\rm cr}$ as a function of $r$, (c) $N_{\rm cr}$ as a function of $p$, and (d) $r_{\rm cr}$ as a function of $N$. When calculating $N_{\rm cr}$ and $r_{\rm cr}$, the simulation values of $T$ were obtained as averages taken over $10^4$ trials. In all figures, the simulations agree well with the theoretical curves. 

From the definition of $N_{\rm cr}$ ($r_{\rm cr}$), $T_{\rm SS}>T_{\rm SW}$ (Strategy SW is advantageous), if $N<N_{\rm cr}$ ($r>r_{\rm cr}$); otherwise $T_{\rm SS}<T_{\rm SW}$ (Strategy SS is advantageous), for fixed $(\alpha,r,p)$ (for fixed $(N,\alpha,p)$). Therefore, these results indicate that Strategy SS is generally advantageous especially for large $N$; however, in limited cases Strategy SW can perform better than Strategy SS, mainly for small $\alpha$, large $r$, large $p$, and small $N$.

\section{Individual (micro-scale) \\ transportation time \\ with the two-lane model}
\label{sec:twolanet}
Now, we introduce the important novel quantity $\tau(n,\alpha,r,p)$. The quantity $\tau$ is defined as the time gap between the start of the simulation and the time when the $n$th-leaving particle leaves the system, which we refer to as the individual transportation time. The quantity $n$ is assumed to be sufficiently small compared to the total number of particles when considering $\tau(n,\alpha,r,p)$. The individual transportation times with Strategy SS and SW are represented as $\tau_{\rm SS}$ and $\tau_{\rm SW}$, respectively, as when we were discussing $T$.

The quantity $\tau_{\rm SS}(n=N)$ clearly coincides with $T_{\rm SS}(N)$ as the first-in-first-out condition is always satisfied with Strategy SS; however, $\tau_{\rm SW}(n=N)$ is generally below $T_{\rm SW}(N)$ since the first-in-first-out condition might not hold with Strategy SW. We note that $\tau_{\rm SW}(n=N)=T_{\rm SW}(N)$ only for $r=0$ and $r=1$, for which the first-in-first-out condition must be satisfied, so $r$ is set to $0<r<1$ in the following discussion.

\subsection{Approximate theoretical analyses of $n_{\rm cr}$}
In this subsection, we theoretically determine approximate values for $n_{\rm cr}$, for which $\tau_{\rm SS}(n=n_{\rm cr})=\tau_{\rm SW}(n=n_{\rm cr})$. 

Unlike we did when approximating $T$, we regard the steady-state flow $Q$ as the average number of particles that pass the right boundary in each step. 

In addition, we assume that (i) $1/\alpha$ time steps are needed on average for the first-entering standing-preference (walking-preference) particle to enter the system~\footnote[1]{Strictly speaking, these required time steps for each lane are different; however, they are basically small unless $r$ approaches very near 0 or 1. Therefore, we simply assume that they are equal to make the calculations easier.}, and that (ii) particles leave the system in the steady state after the first-leaving particle leaves the system.

Furthermore, the number of particles is assumed to be sufficiently large. Especially, the number of all walking-preference particles needs to be larger than the average number of walking-preference particles that leave the system until the time at which the first-leaving standing-preference particle leaves the system.

We next define the threshold $n=N_2$; specifically, the $N_2$th-leaving walking-preference particle leaves the system approximately at the same time when the first-leaving standing-preference particle leaves. Therefore, $N_2$ satisfies
\begin{equation}
(N_2-1)\times\displaystyle\frac{1}{Q_{\rm W}}\approx\displaystyle\frac{1}{\alpha}+\displaystyle\frac{L}{1}-\displaystyle\frac{1}{\alpha}-\displaystyle\frac{L}{1+p},
\end{equation}
which reduces to
\begin{equation}
N_2\approx\frac{r\alpha}{1+r\alpha}\frac{pL}{1+p}+1.
\label{eq:N2}
\end{equation}

Based on the above assumptions, the system behavior can be in three states;  (i) both lanes are not yet in steady state, (ii) only the walking lane is in steady state, and (iii) both lanes are in steady state. Table \ref{tab:state} summarizes the states (flows) of the system and relates them to time steps in terms of the model parameters. 

\setcounter{table}{3}
\renewcommand{\arraystretch}{2}
\begin{table}[htbp]
\centering
\caption{States (flows) of the system at ranges of time steps. `Stand,' `Walk,' and `Entire' represent the standing lane, walking lane, and entire system, respectively. In addition, the notations `R.P.' and `S.S.' stand relaxation process and steady state, respectively.}
\label{tab:state}
\begin{tabular}{c|c|c|c|c}
No. &Time &  Stand & Walk  & Entire \\ \hline \hline
1&$0\leq t<\displaystyle\frac{1}{\alpha}+\displaystyle\frac{L}{1+p}$ & R.P. (0)  & R.P. (0) & R.P. (0) \\ \hline
2&$\displaystyle\frac{1}{\alpha}+\displaystyle\frac{L}{1+p}\leq t<\displaystyle\frac{1}{\alpha}+\displaystyle\frac{L}{1}$ & R.P. (0)  & S.S. $\left(Q_{\rm W}\right)$ & R.P.  $\left(Q_{\rm W}\right)$ \\ \hline
3&$\displaystyle\frac{1}{\alpha}+\displaystyle\frac{L}{1}\leq t$  & S.S. $\left(Q_{\rm S}\right)$  & S.S. $\left(Q_{\rm W}\right)$ & S.S. $\left(Q_{\rm SW}\right)$ \\ \hline
\end{tabular}
\end{table}
\renewcommand{\arraystretch}{1}
\renewcommand{\thefigure}{\arabic{figure}}

Consequently, if we note the first $N_2$ particles leave the system with $Q_{\rm W}$ and the next ($n-N_2$) particles leave with $Q_{\rm SW}$, $\tau_{\rm SW}$ reduces to 
\begin{eqnarray}
\begin{split}
&\tau_{\rm SW}(n,\alpha,r,p)\\
&\approx \left\{ \begin{array}{ll}
\displaystyle\frac{1}{\alpha}+\displaystyle\frac{L}{1+p}+\displaystyle\frac{n-1}{\frac{r\alpha}{1+r\alpha}} & {\rm for} \ n\leq N_2,\\
\\
\displaystyle\frac{1}{\alpha}+\displaystyle\frac{L}{1}+\displaystyle\frac{n-N_2}{\frac{r\alpha}{1+r\alpha}+\frac{(1-r)\alpha}{1+(1-r)\alpha}} & {\rm for} \ n> N_2,
\end{array} \right.
\end{split}
\label{eq:tSW}
\end{eqnarray}
from Eqs. (\ref{eq:QW}) and (\ref{eq:QSW}).

On the other hand, since $\tau_{\rm SS}(n=N)=T_{\rm SS}(N)$, $\tau_{\rm SS}(n,\alpha,p)$ can be written as

\begin{equation}
\tau_{\rm SS}(n,\alpha,p)=T_{\rm SS}(N=n)\approx \frac{L}{1}+\frac{n}{\alpha},
\label{eq:tauSS}
\end{equation}
from Eq. (\ref{eq:TSS}). 

From the definition of $n_{\rm cr}$, and Eqs. (\ref{eq:tSW}) and (\ref{eq:tauSS}), $n_{\rm cr}$ satisfies the following equation:

\begin{equation}
\begin{split}
&\tau_{\rm SS}(n=n_{\rm cr})= \tau_{\rm SW}(n=n_{\rm cr})\\
&\Leftrightarrow \frac{L}{1}+\frac{n_{\rm cr}}{\alpha}=\displaystyle\frac{1}{\alpha}+\frac{L}{1}+\displaystyle\frac{n_{\rm cr}-N_2}{\frac{r\alpha}{1+r\alpha}+\frac{(1-r)\alpha}{1+(1-r)\alpha}}.
\label{eq:ncr1}
\end{split}
\end{equation}
The quantity $\tau_{\rm SW}$ is clearly smaller than $\tau_{\rm SS}$ for $n<N_2$, so $n_{\rm cr}>N_2$. Therefore, $n_{\rm cr}(\alpha,r,p)$ finally reduces to

\begin{equation}
n_{\rm cr}(\alpha,r,p)=\frac{pLr\alpha\{1+(1-r)\alpha\}}{[\{(1-r)^2+r^2\}\alpha+r(1-r)\alpha^2](1+p)}+1,
\label{eq:ncr2}
\end{equation}
using Eqs. (\ref{eq:N2}) and (\ref{eq:ncr1}).

\subsection{Comparison with simulation results}
In this subsection, we compare the theoretical approximations, obtained in the previous subsection, with simulation results. 

\begin{figure}[htbp]
\begin{center}
\includegraphics[width=8.5cm,clip]{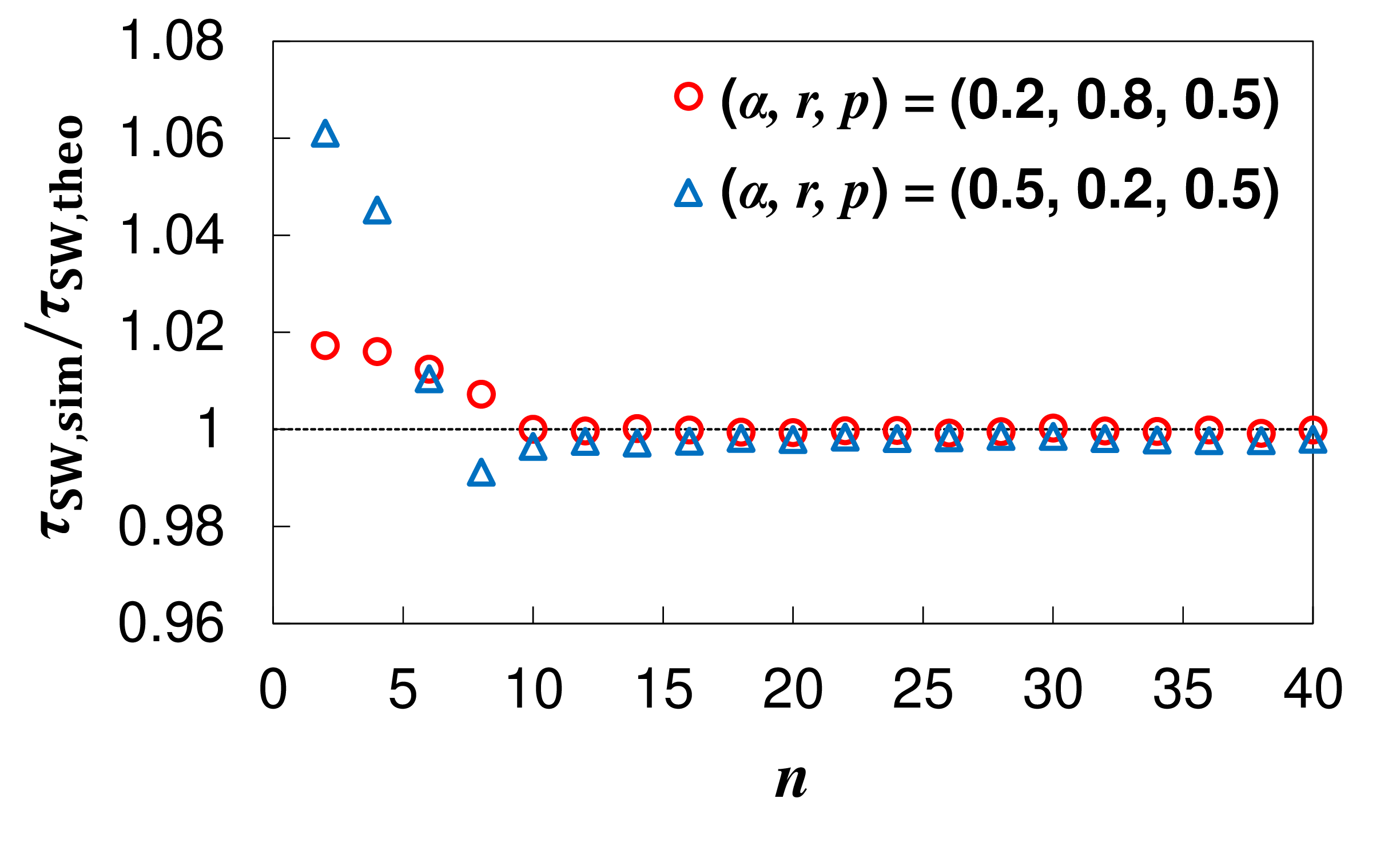}
\caption{(Color Online) Calculated values of the ratio $\tau_{\rm SW,sim}/\tau_{\rm SW,theo}$ as functions of $n$ for $(\alpha,r,p)\in\{(0.2,0.8,1)  {\rm (red \ circles)}, (0.5,0.5,0.5)  {\rm (blue \ triangles)}\}$, with fixed $N=1000$. Both plots begin at $n=2$.}
\label{fig:validityt2d}
\end{center}
\end{figure}

First, Fig. \ref{fig:validityt2d} shows the values of the ratio $\tau_{\rm SW,sim}/\tau_{\rm SW,theo}$ as a function of $n$ for $(\alpha,r,p)\in\{(0.2,0.8,0.5) , (0.5,0.5,0.5)\}$. The simulation values show very good agreement with the theoretical ones for $n>10$. We note that the comparison regarding $\tau_{\rm SS}$ is abbreviated since $\tau_{\rm SS}(n=N)=T_{\rm SS}(N)$.

\begin{figure}[htbp]
\begin{center}
\includegraphics[width=8.5cm,clip]{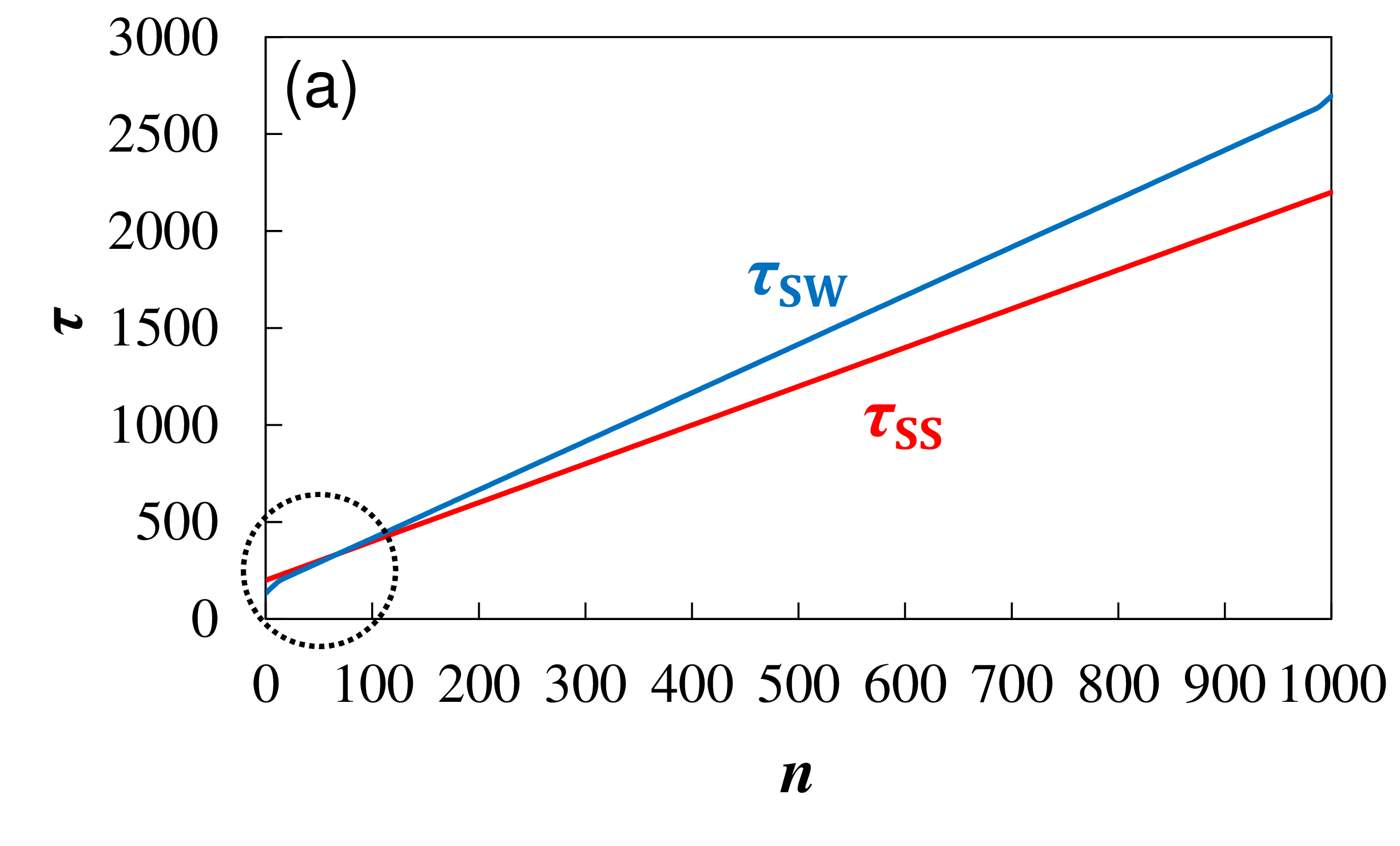}\\
\includegraphics[width=8.5cm,clip]{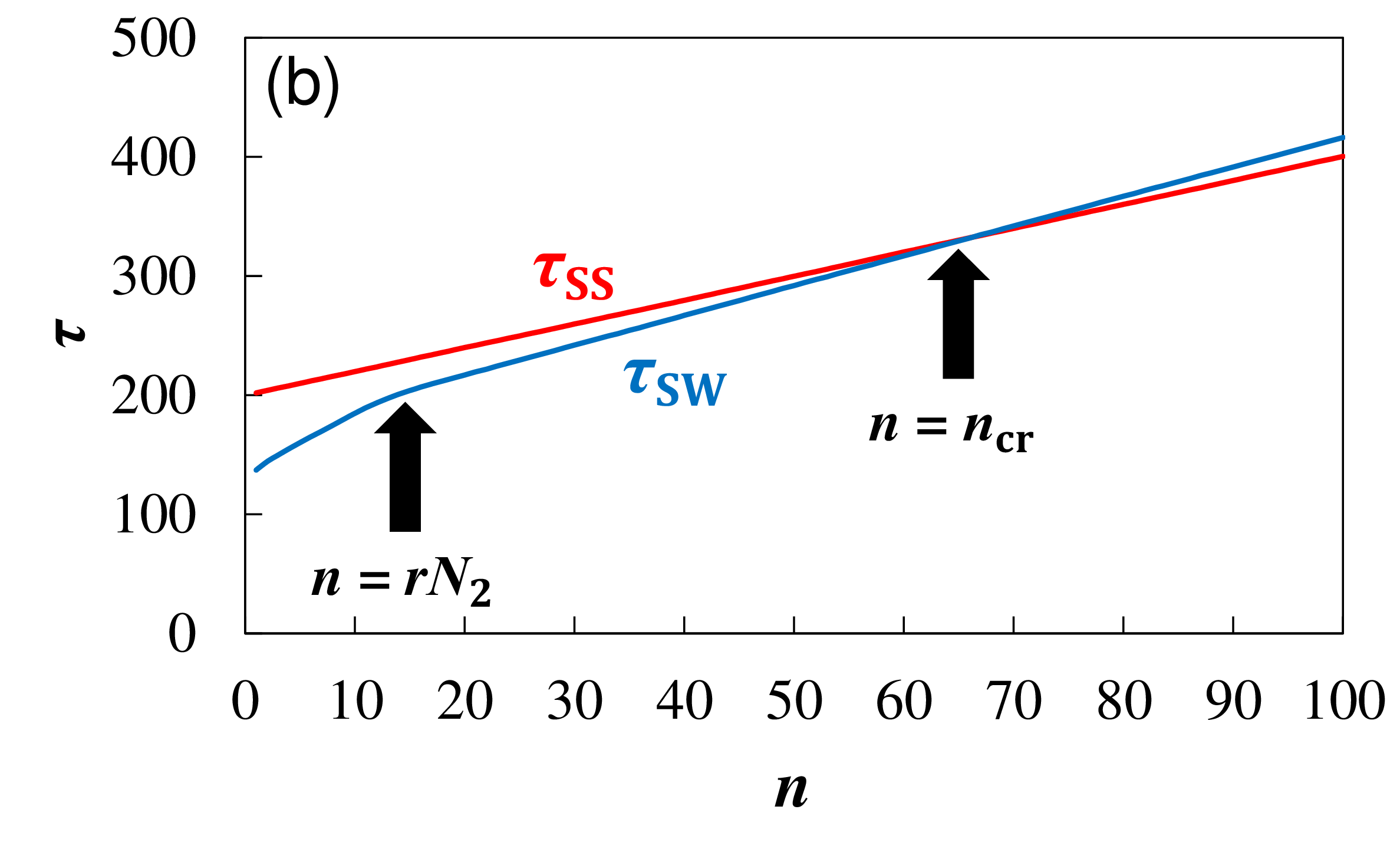}
\caption{(Color Online) (a) Simulation values of $\tau_{\rm SS}$ (red) and $\tau_{\rm SW}$ (blue) as functions of $n$, with fixed $\alpha=0.5$, $r=0.5$, and $p=0.5$. (b) Zoomed-in inset Fig. \ref{fig:n} (a), focusing on the range enclosed by a black dotted circle in Fig. \ref{fig:n} (a). The vicinities of $n=rN_2$ and $n=n_{\rm cr}$ are marked with black arrows, respectively.}
\label{fig:n}
\end{center}
\end{figure}

Next, Fig. \ref{fig:n} compares the simulation values of $\tau_{\rm SS}$ and $\tau_{\rm SW}$ as functions of $n$, with fixed $N=1000$.
In Figs. \ref{fig:n} (a) and (b), we see a point $\tau_{\rm SS}(n)=\tau_{\rm SW}(n)$, in which $n$ is defined as $n=n_{\rm cr}$. This phenomenon can be explained qualitatively as follows. 

At first, some of the first entering waking-preference particles leave the system when Strategy SW is simulated more quickly than when Strategy SS is simulated. In this case, for small $n$ the positive effect of walking has the dominant effect on $\tau_{\rm SW}$, so $\tau_{\rm SS}>\tau_{\rm SW}$. After some time has elapsed, the negative effect of preference becomes dominant, resulting in $\tau_{\rm SS}<\tau_{\rm SW}$.

\begin{figure}[htbp]
\begin{center}
\includegraphics[width=8.5cm,clip]{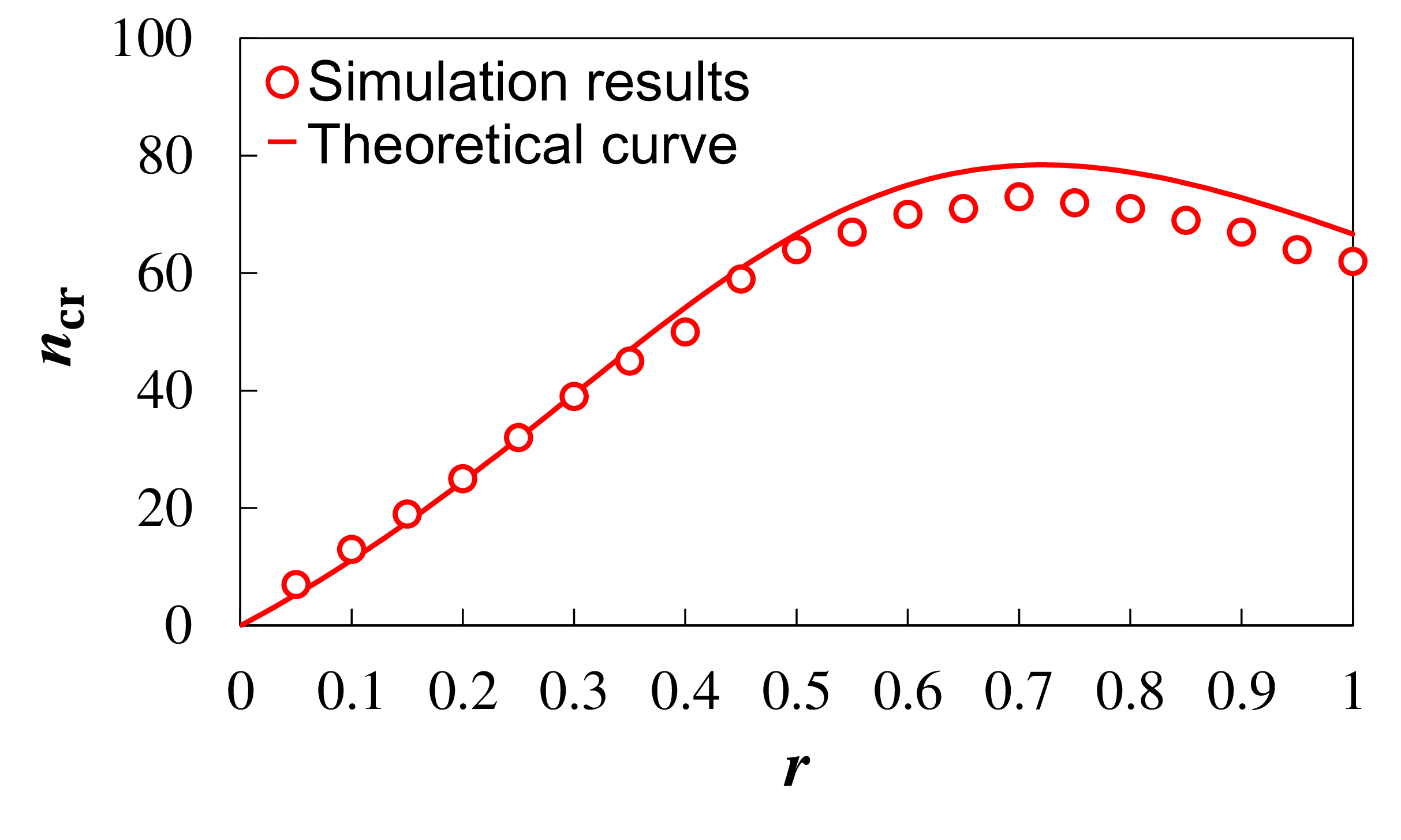}
\caption{(Color Online) Simulation (dots) and theoretical (curve) values of $n_{\rm cr}$ as a function of $r$. The other parameters are fixed as $\alpha=0.5$ and $p=0.5$. We note that the simulation values of $n_{\rm cr}$ satisfy $\tau_{\rm SS}(n=n_{\rm cr})>\tau_{\rm SW}(n=n_{\rm cr})$ and $\tau_{\rm SS}(n=n_{\rm cr}+1)<\tau_{\rm SW}(n=n_{\rm cr}+1)$.}
\label{fig:ncrtau}
\end{center}
\end{figure}

Finally, Fig. \ref{fig:ncrtau} compares the simulation (dots) and theoretical (curve) values for $n_{\rm cr}$ as a function of $r$. In Fig. \ref{fig:ncrtau}, the simulation values show relatively good agreement with the theoretical curves. The upper (lower) bound of the curve of $n_{\rm cr}$ indicates the number of particles that can leave the system more quickly with Strategy SS (Strategy SW) than with Strategy SW (Strategy SS), even if $N$ is large.

In addition, $n_{\rm cr}$ takes the maximal value around $r=0.7$ in Fig. \ref{fig:ncrtau}. If we notice the third term of $\tau_{\rm SW}$ in Eq. (\ref{eq:tSW}) for $n>N_2$, the numerator is maximized when $r=1$ (see also Eq. (\ref{eq:N2})), while the denominator is maximized when $r=0.5$ (see Appendix \ref{sec:QSW}). The former (latter) corresponds to the positive effect of walking (the negative effect of preference). Therefore, for fixed $(n,\alpha,p)$, $\tau_{\rm SW}$ takes its maximum value in the range $0.5<r<1$, which implies that $n_{\rm cr}$ also takes its maximal value in the range $0.5<r<1$.

\section{CONCLUSION}
\label{sec:conclusion}
We have analyzed three strategies for movement on an escalator: (i) two standing lanes (Strategy SS), (ii) one standing lane plus one walking lane (Strategy SW), and (iii) two walking lanes (Strategy WW). These strategies were modeled with a modified two-lane TASEP. The specific contributions of this study are as follows. 

In Sec. \ref{sec:onelane}, we found that the one-lane model with our modified updating rules exhibits only the LD phase for all values of $p$, and that the steady-state flow for any $p$ is identical to that of the original open-boundary TASEP with $p=1$. This indicates that walking on an escalator does not affect steady-state flow, which is consistent with the results in~\cite{yue2018cellular}.

In Sec. \ref{sec:twolaneT}, we considered the total transportation time $T$ using theoretical analysis and numerical simulations. In steady state, $Q_{\rm SS}$ is equal to $Q_{\rm WW}$ and is always greater than $Q_{\rm SW}$. For example, $Q_{\rm SW}$ decreases at $33\%$---$50\%$ compared to $Q_{\rm SS}$ in the most-congested situations ($\alpha=1$). 
This finding is consistent with the results of a pilot performed in the London underground in 2015~\cite{kukadia2016pilot}; when `standing only' was specified for escalators at Holborn station, about $30\%$ more customers could use an escalator during the busiest times when both lanes were used only for standing.

Since $Q_{\rm SW}=Q_{\rm WW}$, counter-intuitively, $T_{\rm SS}\approx T_{\rm WW}$ for sufficiently large $N$, and $T_{\rm SS}$ is generally smaller than $T_{\rm SW}$, indicating that Strategy SS is advantageous for large $N$.
On the other hand, in limited cases for small $N$, Strategy SW can be more useful. Those differences arise because the negative effect of preference generally exceeds the positive effect of walking with Strategy SW for large $N$, while this relation is reversed for small $N$. 

We have determined the reversal point $N=N_{\rm cr}$ and $r=r_{\rm cr}$, which satisfies $T_{\rm SS}(N=N_{\rm cr})=T_{\rm SW}(N=N_{\rm cr})$ and $T_{\rm SS}(r=r_{\rm cr})=T_{\rm SW}(r=r_{\rm cr})$. These values confirm that Strategy SW is advantageous mainly for small $\alpha$, large $r$, large $p$, and small $N$.

In Sec. \ref{sec:twolanet}, we considered the individual transportation time $\tau$ using theoretical analysis and numerical simulations. Even if $N$ is large, in which Strategy SS is advantageous in terms of reducing $T$, the first-leaving $n_{\rm cr}$ particles, mainly including walking-preference particles, can leave the system more quickly under Strategy SW than under Strategy SS. This behavior arises because the positive effect of walking is more active than the negative effect of preference in this case.

Applying the results of the present paper to real situations, we find that encouraging pedestrians to only stand on an escalator, which has recently been suggested by practitioners, is indeed beneficial for improving pedestrian flow, especially when a large number of pedestrians (large $N$) are using a facility. Conversely, providing a walking lane can improve pedestrian flow in limited cases in which the entrance is relatively uncongested (small $\alpha$), the fraction of walking-preference pedestrians is high (large $r$), and the walking velocity is large (large $p$) with a small number of pedestrians (small $N$). In addition, the first-entering walking-preference pedestrians tend to benefit from Strategy SW, even if the number of pedestrians is large (large $N$).

More realistic models and comparisons with field data will be needed to more clearly relate our findings to the real world. For example, combining multiple models of escalators and floor fields, like a cellular-automaton pedestrian model~\cite{burstedde2001simulation}, remains as a future work, along with fitting using actual data. Even so, the simple model discussed above offers useful insights as a first step.

\section*{ACKNOWLEDGMENTS}
We greatly appreciate Airi Goto for her fruitful comments.
This work was partially supported by JST-Mirai Program Grant Number
JPMJMI17D4, Japan, JSPS KAKENHI Grant Number JP15K17583, and MEXT as
gPost-K Computer Exploratory Challengesh (Exploratory Challenge 2:
Construction of Models for Interaction Among Multiple Socioeconomic
Phenomena, Model Development and its Applications for Enabling Robust
and Optimized Social Transportation Systems) (Project ID: hp180188).

\appendix

\begin{table*}
\flushleft
\begin{eqnarray*}
\begin{split}
T_{\rm SW}&\approx\displaystyle\sum_{k=1}^{N+1} P(k)t_{\rm SW}(k)
\\
&=\displaystyle\frac{1}{\alpha}+\displaystyle\sum_{k=1}^{N_1} \{P(k)\times(T_{\rm SW}^{\rm s}(N-k+1)+T_{\rm S}^{\rm f})\} + \left(1-\sum_{k=1}^{N_1} P(k)\right)\times(T_{\rm SW}^{\rm s}(N)+T_{\rm W}^{\rm f}) 
\\
&=\displaystyle\frac{1}{\alpha}+\displaystyle\sum_{k=1}^{N_1} \left\{(1-r)r^{k-1}\left(\frac{N-k}{Q_{\rm SW}}+\frac{L}{1}\right)\right\} + \left(1-\sum_{k=1}^{N_1} (1-r)r^{k-1}\right)\times\left(\frac{N-1}{Q_{\rm SW}}+\frac{L}{1+p}\right)
\\
&=\displaystyle\frac{1}{\alpha}+\frac{N-1}{Q_{\rm SW}}-\displaystyle\sum_{k=1}^{N_1}\frac{(k-1)(1-r)r^{k-1}}{Q_{\rm SW}}+\frac{(1-r^{N_1})L}{1}+\displaystyle\frac{r^{N_1}L}{1+p} \ \ \ \ \ \ \ \ \ \ \ \ \ \ \ \ \ \ \ \ \ \ \ \ \ \ \ \ \ \ \ \ \ \ \ \ \ \ \ \ \ \ \ \ \ \ \ \ \ \ \ \ \ {\rm (B1)} 
\\
&=\left\{ \begin{array}{ll}
\displaystyle\frac{1}{\alpha}+\displaystyle\frac{N-1}{\frac{r\alpha}{1+r\alpha}+\frac{(1-r)\alpha}{1+(1-r)\alpha}}-\displaystyle\frac{\frac{r}{1-r}\{1-N_1 r^{N_1-1}+(N_1-1)r^{N_1}\}}{\frac{r\alpha}{1+r\alpha}+\frac{(1-r)\alpha}{1+(1-r)\alpha}}+\displaystyle\frac{(1-r^{N_1})L}{1}+\displaystyle\frac{r^{N_1}L}{1+p} &\ \  {\rm for} \ r\neq 1
\\
\\
\displaystyle\frac{1}{\alpha}+\displaystyle\frac{N-1}{\frac{\alpha}{1+\alpha}}+\frac{L}{1+p} & \ \  {\rm for} \ r= 1\\
\end{array} \right.
\\
&=\displaystyle\frac{1}{\alpha}+\displaystyle\frac{N-1}{\frac{r\alpha}{1+r\alpha}+\frac{(1-r)\alpha}{1+(1-r)\alpha}}-\displaystyle\frac{\sum_{k=1}^{N_1}(1-r^{N_1-k})r^{k}}{\frac{r\alpha}{1+r\alpha}+\frac{(1-r)\alpha}{1+(1-r)\alpha}}+\displaystyle\frac{(1-r^{N_1})L}{1}+\displaystyle\frac{r^{N_1}L}{1+p}
\end{split}
\end{eqnarray*}
\end{table*}

\section{Properties of $Q_{\rm SW}$}
\label{sec:QSW}
In this appendix, we briefly give the details of the properties of $Q_{\rm SW}$, i.e., Eq. (\ref{eq:QSW}), as a function of $\alpha$ and $r$. 

Defining $f(\alpha,r)$ as 
\begin{equation}
Q_{\rm SW}=f(\alpha,r)=\frac{(1-r)\alpha}{1+(1-r)\alpha}+\frac{r\alpha}{1+r\alpha},
\end{equation}
$f(\alpha,r)$ can be rewritten as
\begin{equation}
f(\alpha,r)=2-\frac{1}{1+r\alpha}-\frac{1}{1+(1-r)\alpha}.
\end{equation}
Therefore, for fixed $r$ ($0\leq r\leq1$), $f(\alpha,r)=f(\alpha)$ is obviously a monotonically increasing function of $\alpha$ ($0\leq \alpha\leq1$). We assume that $\alpha$ is constant hereafter.

Replacing $r\alpha$ with $x$ for simplicity, $f(\alpha,r)=f(x)$ can be written as 
\begin{equation}
f(x)=\frac{\alpha-x}{1+\alpha-x}+\frac{x}{1+x},
\end{equation}
where $0\leq x \leq \alpha$.
Taking the first derivation of $f(x)$, $df(x)/dx$ is calculated as
\begin{equation}
\frac{df(x)}{dx}=\frac{(\alpha-2x)\{2+\alpha+4x(\alpha-x)\}}{(1+x)^2(1+\alpha-x)^2}.
\end{equation}

The results of the first derivative test are summarized in Tab. \ref{tab:derivation}. $f(x)$ takes its maximum value $2\alpha/(2+\alpha)$ with $x=\alpha/2$ ($r=0.5$). 

\renewcommand{\arraystretch}{1.6}
\begin{table}[htbp]
\centering
\caption{Result of the first derivative test of $f(x)$}
\label{tab:derivation}
\begin{tabular}{c|c|c|c|c|c}
$x$ & 0 &... & $\displaystyle\frac{\alpha}{2}$ &...&$\alpha$ \\ \hline 
$f(x)$ & &$\nearrow$ & $f\left(\displaystyle\frac{\alpha}{2}\right)$ &$\searrow$ & \\ \hline
$\displaystyle\frac{df(x)}{dx}$ & $+$ & $+$ & 0 & $-$ & $-$\\ 
\end{tabular}
\end{table}
\renewcommand{\arraystretch}{1}

\begin{figure}[htbp]
\begin{center}
\includegraphics[width=7cm,clip]{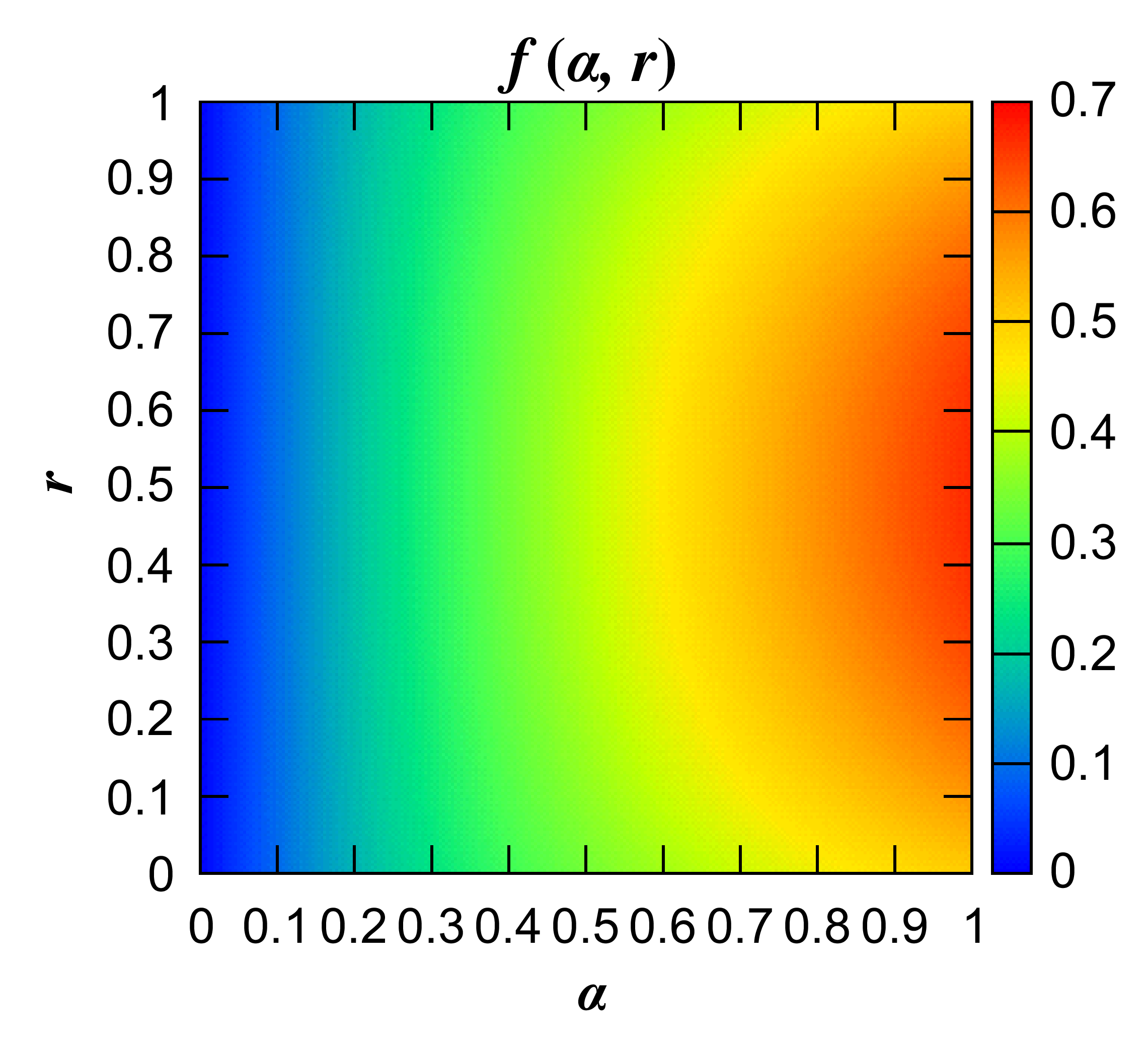}
\caption{(Color Online) Theoretical values of $f(\alpha,r)$ for various $(\alpha,r)$.}
\label{fig:fx}
\end{center}
\end{figure}

Therefore, $f(\alpha,r)$ takes its maximum value ($f(\alpha,r)=2/3$) when $\alpha=1$ and $r=0.5$.
Figure \ref{fig:fx} shows the calculated values of $f(\alpha,r)$ in the $(\alpha,r)$ plane. From Fig. \ref{fig:fx}, $f(\alpha,r)$ is sensitive to $r$ for large $\alpha$.

\section{Detailed calculation of $T_{\rm SW}$}
\label{sec:TSW}

In this appendix, we discuss the detailed calculations for obtaining the general form of $T_{\rm SW}$. 

The specific calculations are summarized as Eq. (B1). We note that the third term of the fourth line in Eq. (B1) is the summation of an arithmetico-geometric sequence, and therefore, we can obtain a closed-form expression.

\section{Discussion of the behavior of $T_{\rm SW}-T_{\rm SS}$ \\ when $\alpha_1<\alpha\leq1$ for sufficiently large $N$}
\label{sec:monotonicity}
In this appendix, we briefly discuss the behavior of $T_{\rm SW}-T_{\rm SS}$ when $\alpha_1<\alpha\leq1$ for sufficiently large $N$. 

For sufficiently large $N$; especially $N\gg N_0$, $N_1=N_0$, due to $N_1=\min(N_0,N)$. Therefore, using Eq. (\ref{eq:TSWTSS}), $T_{\rm SW}-T_{\rm SS}$ can be represented as follows:
\begin{equation}
\begin{split}
&T_{\rm SW}-T_{\rm SS}\\
&=(N-1)\left(\frac{1}{Q_{\rm SW}}-\frac{1}{Q_{\rm SS}}\right)-\displaystyle\frac{\sum_{k=1}^{N_0}(1-r^{N_0-k})r^{k}}{Q_{\rm SW}}\\
&\quad-\displaystyle\frac{pr^{N_1}}{1+p}\frac{L}{1}\\
&=\displaystyle\frac{A}{\frac{1-r}{1+(1-r)\alpha}+\frac{r}{1+r\alpha}}-\displaystyle\frac{B}{2-\frac{1}{1+(1-r)\alpha}-\frac{1}{1+r\alpha}}-C,
\end{split}
\label{eq:monotonicity}
\end{equation}
where $A$, $B$, and $C$ can be regard as constant values; specifically,
\begin{equation}
A=\{1-r(1-r)\}(N-1)>0,
\end{equation}
\begin{equation}
B=\sum_{k=1}^{N_0}(1-r^{N_0-k})r^{k}>0,
\end{equation}
and
\begin{equation}
C=\displaystyle\frac{pr^{N_0}}{1+p}\frac{L}{1}>0.
\label{eq:C}
\end{equation}

Noting that the the denominators of the first (second) terms of Eq. (\ref{eq:monotonicity}) clearly increase (decrease) monotonically with respect to $\alpha$, from Eq. (\ref{eq:monotonicity})---(\ref{eq:C}) we see that $T_{\rm SW}-T_{\rm SS}$ is a monotonically increasing function of $\alpha$.

Considering that $T_{\rm SW}-T_{\rm SS}>0$ for $\alpha=\alpha_1$ and sufficiently large $N$, $T_{\rm SW}-T_{\rm SS}$ always exceeds 0 for $\alpha_1<\alpha\leq1$.

\section{Discussion of the behavior of $g(\alpha)$ \\ when $0<\alpha\leq\alpha_1$ for sufficiently large $N$}
\label{sec:galpha}
Here, we discuss the details of the behavior of $g(\alpha)$ when $0<\alpha\leq\alpha_1$ for sufficiently large $N$. The function $g(\alpha)$ can be rewritten as follows:
\begin{equation}
\begin{split}
g(\alpha)&=1-\frac{1-r}{1+(1-r)\alpha}-\frac{r}{1+r\alpha}\\
&\quad-\frac{rpL}{(1+p)(N-1)}\left\{\frac{1-r}{1+(1-r)}+\frac{r\alpha}{1+r\alpha}\right\}\\
&=1-\frac{2rpL}{(1+p)(N-1)}-\frac{1-r-\frac{rpL}{(1+p)(N-1)}}{1+(1-r)\alpha}\\
&\quad-\frac{r-\frac{rpL}{(1+p)(N-1)}}{1+r\alpha}\\
&=1-\frac{2rpL}{(1+p)(N-1)}-\frac{D}{1+(1-r)\alpha}-\frac{E}{1+r\alpha},
\end{split}
\end{equation}
where $D$ and $E$ are represented as 
\begin{equation}
D=1-r-\frac{rpL}{(1+p)(N-1)}
\end{equation}
and
\begin{equation}
E=r-\frac{rpL}{(1+p)(N-1)}.
\end{equation}

For sufficiently large $N$, which especially satisfies $D>0$ and $E>0$; specifically, 
\begin{equation}
N>\frac{pL}{1+p}\max\left(\frac{r}{1-r},1\right)+1,
\end{equation}
$dg(\alpha)/d\alpha$ can be calculated as
\begin{equation}
\frac{dg(\alpha)}{d\alpha}=\frac{(1-r)D}{\{1+(1-r)\alpha\}^2}+\frac{rE}{(1+r\alpha)^2}>0.
\end{equation}
Therefore, for sufficiently large $N$, $g(\alpha)$ is a monotonically increasing function when $0<\alpha\leq\alpha_1$. Considering this fact together with $g(0)=0$, $g(\alpha)>0$, and therefore,
\begin{equation}
T_{\rm SW}-T_{\rm SS}>0,
\end{equation}
when $0<\alpha\leq\alpha_1$.

\end{document}